\documentclass[9pt,twocolumn,twoside]{osajnl}
\journal{pr} 
\setboolean{shortarticle}{false}
\usepackage{lineno}
\usepackage{graphicx} 
\usepackage{subfigure}

\title{A round-trip multi-band quantum access network}

\author[1]{Yuehan Xu}
\author[1,2,3,*]{Tao Wang}
\author[1]{Huanxi Zhao}
\author[1,2,3]{Peng Huang}
\author[1,2,3,$\dag$]{Guihua Zeng}

\affil[1]{State Key Laboratory of Advanced Optical Communication Systems and Networks, Center of Quantum Sensing and Information Processing, Shanghai Jiao Tong University, Shanghai 200240, China}
\affil[2]{Shanghai Research Center for Quantum Sciences, Shanghai 201315, China}
\affil[3]{Hefei National Laboratory, CAS Center for Excellence in Quantum Information and Quantum Physics, Hefei, Anhui 230026, China}
\affil[*]{tonystar@sjtu.edu.cn}
\affil[$\dag$]{ghzeng@sjtu.edu.cn}

\begin{abstract}

The quantum network makes use of the quantum states to transmit data, which will revolutionize classical communication and allow for some breakthrough applications. The quantum key distribution (QKD) is one prominent application of quantum networks, and can protect the data transmission through quantum mechanics. In this work, we propose an expandable and cost-effective quantum access network, in which the round-trip structure makes quantum states travel in a circle to carry the information, and the multi-band technique is proposed to support multi-user access. Based on the round-trip multi-band quantum access network, we realize multi-user secure key sharing through the continuous-variable QKD (CV-QKD) protocol. Due to the encoding characteristics of CV-QKD, the quadrature components in different frequency bands can be used to transmit key information for different users. The feasibility of this scheme is confirmed by comprehensive noise analysis, and is verified by a proof-of-principle experiment. The results show that each user can achieve excess noise suppression and $600\;\rm{bps}$ level secure key generation under $30\;\rm{km}$ standard fiber transmission. Such networks have the ability of multi-user access theoretically and could be expanded by plugging in simple modules. Therefore, it paves the way for near-term large-scale quantum secure networks.

\end{abstract}

\setboolean{displaycopyright}{false}

\begin{document}	
\maketitle
\fancyhead{} %Clear header
%\fancyfoot{} %Clear footer
%\fancyhf{} %Clear header and footer
	
\section*{Introduction}

\indent Quantum network is an interconnected network that makes use of the properties of quantum to transmit data, and could revolutionize the way of information exchange in the future \cite{khatri2021towards}. The quantum network \cite{kimble2008quantum} has many breakthrough applications, such as the quantum secure network through quantum key distribution (QKD) \cite{dianati2008architecture,stucki2011long,wang2014field,bedington2017progress,tajima2017quantum,kiktenko2017demonstration,zhang2018large,PRXQuantum.3.020341} or quantum secure direct communication \cite{wengerowsky2018entanglement,qi202115}. In addition, it could also perform other tasks impossible in classical physics, such as distributed quantum computing \cite{simon2017towards}, and accurate global timing \cite{komar2014quantum} that will bring us accurate navigation and earth sensing. Moreover, the quantum network could even lead to accurate telescopes \cite{gottesman2012longer} and new fundamental tests of quantum nonlocality, quantum teleportation \cite{bouwmeester1997experimental}, and quantum gravity \cite{rideout2012fundamental}.

\indent QKD is one prominent application of the quantum networks, which is the core technology of the quantum secure communication \cite{pirandola2020advances}. It can provide secure keys for legal parties even in the presence of eavesdroppers, and its theoretical security is guaranteed by the basic principles of quantum mechanics \cite{gisin2002quantum}. The QKD protocol can be divided into discrete-variable QKD (DV-QKD) \cite{bennett2020quantum} and continuous-variable QKD (CV-QKD) \cite{polkinghorne1999continuous,ralph2000security,grosshans2002continuous,grosshans2002reverse,grosshans2003quantum,weedbrook2004quantum,grosshans2004continuous,navascues2006optimality,garcia2006unconditional,lodewyck2007tight,sudjana2007tight,renner2009finetti,christandl2009postselection,leverrier2009unconditional,zhao2009asymptotic,leverrier2010simple,leverrier2010finite,leverrier2011continuous,leverrier2013security,leverrier2017security,bradler2018security,ghorai2019asymptotic} according to the physical quantity that carries key information. According to the quantum technology roadmap released by OIDA \cite{counts2020oida}, the point-to-point QKD has been mature to build network and gradually commercialized \cite{lance2005no,qi2007experimental,jouguet2013experimental,qi2015generating,soh2015self,qu2016rf,huang2016long,huang2016field,kleis2017continuous,laudenbach2019pilot,zhang2019integrated,zhang2019continuous,zhang2020long,wang2020high,ren2021demonstration,liu2021homodyne}. A representative quantum secure network is the Beijing-Shanghai trunk line \cite{chen2021integrated}, which achieves a quite long distance transmission. In addition, Cambridge quantum metropolitan area network is constructed with high bandwidth data transmission \cite{dynes2019cambridge}. Furthermore, quantum networks in the United Kingdom have been operating for several years with 3 nodes separated by 5-10 km optical fiber \cite{wonfor2021quantum}. The 46-node quantum metropolitan area network in Hefei realizes real-time voice telephone, text messaging and file transmission \cite{chen2021implementation}. For multi-user access, the quantum access network proposed by Bernd Frohlich et al. \cite{frohlich2013quantum} is the first scheme to realize the upstream quantum access network between users and a common node.

\indent The physical implementation of a quantum network for QKD is an important issue. On the one hand, constructors need to consider the coverage of the quantum network, which can be divided into backbone networks, metropolitan area networks, and access networks. On the other hand, builders should also be concerned with quantum network topologies, such as star, tree, and mesh. For different coverage and topology, the physical structure of the quantum network is a vital issue. Considering the preparation and measurement of quantum states by multi users, the quantum states generated from different sources require a large number of detectors, while those generated from the same source will interfere with each other and be difficult to be separated. Besides, an alternative scheme is to build large-scale networks based on the mature point-to-point QKD system. However, limited by its complexity, this scheme could not support access services for such a large number of users. Therefore, a new type of quantum network architecture needs to be proposed to simplify the original complex network with multi-user access.

\indent Therefore, we propose an expandable and cost-effective quantum network physical structure, called round-trip multi-band quantum access network (RM-QAN), to improve network performance and support multi-user access. The round-trip multi-band network structure utilizes quantum states travel in a circle to transmit data, and uses the frequency division multiplexing (FDM) technique to isolate different users. In RM-QAN, each user only requires one modulator and circulator when plugging into the network. In addition, only one laser and one detector are needed to build the entire network, which is flexible and cost-effective. Such networks could theoretically support multi-user access without performance penalty. And the network users can achieve $600\;\rm{bps}$ level quantum secure key generation under $30\;\rm{km}$ network range. Moreover, the network scalability and noise suppression are excellent, which is a promising solution for building near-term multi-user quantum secure networks.

\indent In this paper, our RM-QAN scheme is introduced in detail. First, we describe the physical structure of RM-QAN and expound on its advantages. In addition, we use RM-QAN to realize CV-QKD and evaluate the performance of QKD through complete noise analysis. Based on this physical structure, we construct a proof-of-principle experimental platform, and verify the feasibility of multi-user secure key sharing. Finally, we come up with a conclusion.

\section*{Result}

\subsection*{Physical structure of RM-QAN}

\indent Our RM-QAN physical structure is described as follows. In the quantum access network, the quantum network unit (QNU) is the device held by the user, which corresponds to the optical network unit (ONU) \cite{yeh200940,xia2015time,wang2015demonstration} in the optical access network. The QNU does not need to be responsible for receiving light in our scheme, nor does it need to be responsible for generating light sources. Instead, QNU only needs to modulate the key information. It reduces the overall system cost of the quantum access network. The work of generating and receiving light is all done by the quantum line terminal (QLT). QLT is the terminal equipment used to connect quantum trunks, which corresponds to the optical line terminal (OLT) \cite{yeh200940,xia2015time,wang2015demonstration} in the optical access network. The round-trip structure is divided into two stages. Firstly, the optical carrier is transmitted from the QLT side to the QNU side. There is a loss of 1/N in this part, but it is not involved in the transmittance of QKD. Secondly, the optical carrier is transmitted from the QNU side to the QLT side. The light transmitted to the QLT side is brought together from signals of all users. There is also a loss of 1/N in this part, which is involved in the transmittance of QKD. Therefore, the multi-user quantum network can be completed with one laser and one detector, which is more efficient than other schemes. In our scheme, according to the upstream transmission direction, it can be understood that QNU is Alice and QLT is Bob in the classical scheme. 

\indent For a single user, this QKD scheme is actually a plug-and-play QKD architecture. The plug-and-play scheme was proposed in Ref. \cite{muller1997plug} and experimentally verified in DV-QKD \cite{stucki2002quantum,zhao2006experimental,zhang2014reference} and CV-QKD \cite{huang2016continuous}. In addition, the plug-and-play scheme has corresponding commercial QKD systems \cite{Clavis2}. For security, QNU can further randomizes the global phase of each pulse, measures the incoming intensity, and introduces sufficient attenuation, and the standard security proofs of a one-way system can be used \cite{gisin2006trojan}, which has been used in the CV case \cite{huang2016continuous}.

\begin{figure*}
\centering
\includegraphics[width=1\linewidth]{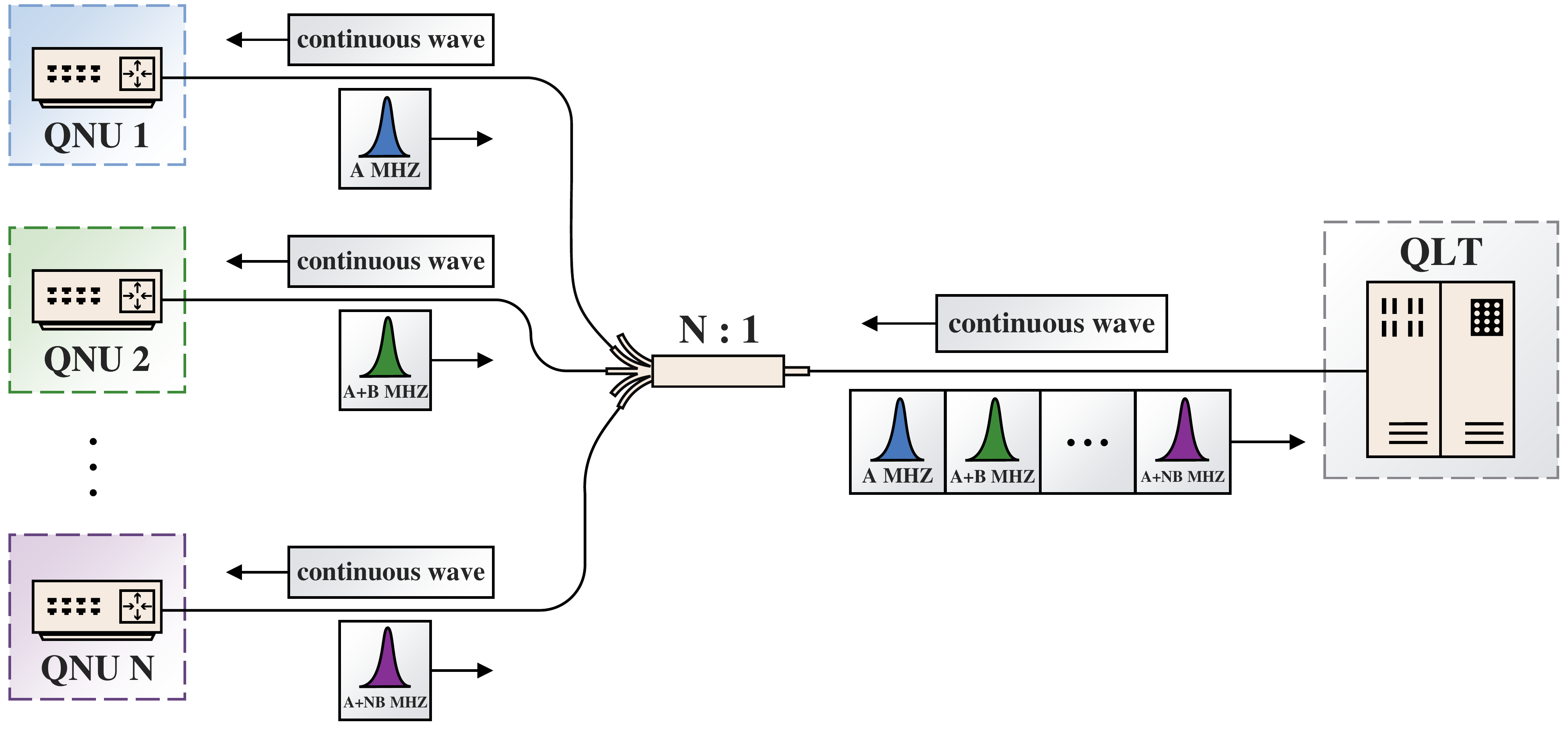}
\caption{Schematic diagram of the round-trip multi-band quantum access network (RM-QAN). First, the continuous wave is generated and transmitted by the quantum line terminal (QLT). Then, the continuous wave is divided into $N$ pieces through an N:1 splitter. The quantum network unit (QNU) will modulate the key information on different carrier frequencies to be distinguished clearly on the spectrum. Then the key is transmitted back through the round-trip structure, and the modulated signal light is returned to the splitter. Finally, the signal light is passed back to the QLT. QLT will demodulate the received signal.}
\label{fig2}
\end{figure*}

\indent FDM is a multiplexing technology that modulates multiple baseband signals to different frequency carriers and then superimposes them to form a composite signal. The FDM method is like a ``frequency modulation (FM) radio system'', which can be tuned to different frequencies to receive information from different sources. In the implementation, since the quadrature components can be modulated on different frequency bands, it can be used to transmit the quantum key \cite{kleis2017continuous,laudenbach2019pilot}, and therefore the FDM method can be adopted for multi-user key distribution. The time division multiplexing (TDM) is an alternative scheme, but it has a high requirement of time slot control. The principle of our scheme is described in detail below.

\indent As can be seen from the schematic diagram of RM-QAN in FIG. \ref{fig2}, there are multiple QNUs corresponding to one QLT. First, the continuous wave is generated and transmitted by QLT in the optical layer. Then, it is divided into $N$ pieces through an N:1 splitter, which does not carry any data. After receiving the light, QNU will conduct secure key modulation through a radio frequency (RF) signal, which carries the information on different frequency bands, and can be distinguished clearly on the spectrum. QNU selects the number $k \in\{0,1,2,3\}$ to form a random sequence of length $n$ with equal probability. Then QNU prepares $n$ coherent states according to the random sequence. The $i$th coherent state can be expressed as
\begin{equation}
	\left|\alpha_{k}\right\rangle=\left|\alpha e^{i(2 k+1) \pi / 4}\right\rangle,i\in\{0,1,\cdots,n\},
\label{fu1}
\end{equation}
where $\alpha^{2}=V_{\mathrm{QNU}}/2=V_{\mathrm{A}}/2$. $V_{\mathrm{QNU}}$ is the modulation variance of QNU in our scheme. Since QNU is equivalent to Alice, all $V_{\mathrm{QNU}}$ below are replaced by the more familiar expression $V_{\mathrm{A}}$. Assuming that the carrier frequency registered for the first QNU is $A\;\rm{MHz}$ and the carrier frequency interval of each adjacent QNU is $B\;\rm{MHz}$, the carrier frequency modulated by the second QNU can be expressed as $A+B\; \rm{MHz}$. Similarly, the carrier frequency modulated by the $N$th QNU is $A+N\cdot B\;\rm{MHz}$. In the modulation process, the original signal $k$ is separated into two groups of information $a$ and $b$ in binary, where $a$ represents the first binary number and $b$ represents the second binary number. Then, the baseband signal is modulated according to the corresponding carrier frequency of each QNU. Therefore, the transmitted signal can be represented by
\begin{equation}
f(t)=a \sin \left(\omega t \right)+b \cos \left(\omega t\right),
\label{ft}
\end{equation}
where $\omega$ is the carrier frequency, and $t$ is the time series. The advantage of this modulation method is that the signal can keep a four-state orthogonal form. Then the key is transmitted back through the round-trip structure, and the modulated signal light is returned to the splitter. The signal light modulated by each QNU is gathered together by the splitter to form a signal light containing $N$ frequency bands. Finally, the signal light is passed back to the QLT through the round-trip structure.

\indent After receiving through a coherent receiver, QLT will get a mixed multi-band spectrum, in which the information is indistinguishable in the time domain but can be clearly distinguished in the frequency domain. QLT first checks all bands that QNU has registered. When accessing the network, each QNU needs to register its frequency band, i.e., carrier frequency. QLT checks the registered bands to see which QNU is currently communicating. This registration method can effectively prevent Eve from using illegal frequency bands to obtain information. For these bands, QLT uses bandpass filtering to separate them. Then, QLT performs the first phase shift recovery for the information, which addresses the optical phase drift during the signal transmission. After that, Since the key obtained at this time still carries out spectrum shifting, QLT needs to remove its carrier by coherent demodulation for each QNU. Coherent demodulation can obtain the baseband signal $g_{a}(t)$ and $g_{b}(t)$. The formula of coherent demodulation is
\begin{equation}
\begin{aligned}
g_{a}(t)=&f(t) \sin (\omega t)=\frac{b}{2} \sin (2 \omega t)-\frac{a}{2} \cos (2 \omega t)+\frac{a}{2}, \\
g_{b}(t)=&f(t) \cos (\omega t)=\frac{a}{2} \sin (2 \omega t)+\frac{b}{2} \cos (2 \omega t)+\frac{b}{2},
\label{gt}
\end{aligned}
\end{equation}
where $\omega$ is the carrier frequency, and $t$ is the time series. Then $g_{a}(t)$ and $g_{b}(t)$ are low-pass filtered to obtain the baseband signal for each QNU. After that, QLT downsamples the obtained data according to the symbol rate. QLT then performs a second phase shift recovery to complete the original data restoration. The second recovery is because the data may have rotation as a whole due to the phase of the RF signal, and the original data needs to be obtained by the inverse operation as a whole. Finally, QLT does the frame synchronization for the data alignment.

\indent To prove that the interference between each user's signal doesn’t exist. Assuming that
\begin{equation}
\begin{aligned}
E_{\rm{S}N}(t)&=A_{\rm{S}N} \cos \left(\omega_{\rm{S}N} t+\phi_{\rm{S}N}\right), \\
E_{\rm{L}}(t)&=A_{\rm{L}} \cos \left(\omega_{\rm{L}} t+\phi_{\rm{L}}\right),
\end{aligned}
\end{equation}
where $E_{\rm{S}N}$ represents the signal of the $N$th user, $E_{\rm{L}}$ represents the local oscillator (LO), $A_{\rm{S}N}$ and $A_{\rm{L}}$ denote the power, $\omega_{\rm{S}N}= \omega_{\rm{O}}+\omega_{N}$ and $\omega_{\rm{L}}= \omega_{\rm{O}}$ are the frequency, $\omega_{\rm{O}}$ represents the frequency of light, $\omega_{N}$ represents the carrier frequency of the $N$th user, $\phi_{\rm{S}N}$ and $\phi_{\rm{L}}$ represent the phase, $t$ denotes the time series.
The output optical power after the coherent detection is
\begin{equation}
\begin{aligned}
P&=K\left(\left|E_{\rm{S} 1}+\cdots+E_{\rm{S} N}+E_{\rm{L}}\right|^2-\left|E_{\rm{S} 1}+\cdots+E_{\rm{S} N}-E_{\rm{L}}\right|^2 \right) \\
&=4 K \sum_{\rm{i=1}}^N \left[ A_{\rm{S} i} A_{\rm{L}} \cos \left(\omega_{\rm{S} i} t+\phi_{\rm{S} i} \right) \cos \left(\omega_{\rm{L}} t+\phi_{\rm{L}}\right)\right] \\
&=2 K \sum_{\rm{i=1}}^N \sqrt{P_{\rm{L}} P_{\rm{S} i}} \left\{\cos \left(\omega_{\rm{i}} t+\phi_{\rm{S} i}-\phi_{\rm{L}}\right)\right.\\
&\left. +\cos\left[\left(2\omega_{\rm{O}}+\omega_{\rm{i}} \right)t+\phi_{\rm{S} i}+\phi_{\rm{L}}\right]\right\},
\end{aligned}
\end{equation}
where $K$ is the coefficient of the photoelectric conversion. Therefore, the beat frequencies between different user’s signals are eliminated after the coherent detection. The frequency of the latter term $\sum_{\rm{i=1}}^N \cos\left[\left(2\omega_{\rm{O}}+\omega_{\rm{i}} \right)t+\phi_{\rm{S} i}+\phi_{\rm{L}}\right]$ is far beyond the bandwidth of the detector and cannot be detected. Only the former term $\sum_{\rm{i=1}}^N \cos \left(\omega_{\rm{i}} t+\phi_{\rm{S} i}-\phi_{\rm{L}}\right)$ still exists. In conclusion, interference between different users doesn’t exist after the coherent detection.

\indent This round-trip structure is mainly vulnerable to Eve's practical security attacks, including the phase remapping attack \cite{xu2010experimental, xu2020secure} and the Trojan-Horse attack \cite{gisin2006trojan}. For phase remapping attack, the effective solutions are that QNU checks the arrival time of the reference pulse and the signal pulse by monitoring, verifying that she is applying the correct modulations to her states \cite{xu2010experimental}. For Trojan-horse attack, since we cannot use the isolator in the two-way structure, the filter can be used to exclude the eavesdropper's input light \cite{gisin2006trojan}. In addition, there are three technical countermeasures: 1. installing a watchdog detector with a switch at the entrance of QLT that randomly routes a small fraction of incoming signals to this detector. 2. opening the door for Eve for a smaller time duration. 3. reducing the width of phase modulation voltage pulse \cite{jain2014trojan}. Certainly, from a theoretical perspective, a higher amount of privacy amplification can help QNU and QLT to destroy the partial information of Eve. We need to estimate the maximum leakage due to Trojan-horse attacks, and incorporate these elements in the security proof \cite{gisin2006trojan,jain2014risk,lucamarini2015practical}.

\indent The round-trip multi-band physical structure has many advantages compared with downstream and upstream quantum access networks. In the downstream quantum access network, the transmitters are located at QLT, while the receivers are located at QNU \cite{huang2021realizing}. This downstream scheme has two major disadvantages. First, each QNU in the network requires a detector, which is normally expensive and difficult to operate. Second, it is impossible to locate the data of each user certainly. Therefore, all detectors must run at the same speed as the transmitter to avoid missing the key, which means that most of the detector's bandwidth is unused. In the upstream quantum access network, QNU is responsible for transmitting the key, and QLT is responsible for receiving the key \cite{frohlich2013quantum}. The upstream scheme still requires multiple laser sources to prepare a quantum signal, which is difficult for ordinary users to afford. In addition, the network capacity has to be determined before building the upstream quantum access network. 

\indent RM-QAN solves the above issues well. In our scheme, each QNU only needs one modulator and circulator to plug into the network, and the entire network only requires one laser and one detector, which is efficient. Moreover, the bandwidth of the detector is fully used due to FDM. In addition, the network scalability and noise suppression are wonderful, which means a large number of users can easily access it at any time.

\subsection*{CV-QKD based on RM-QAN}

\subsubsection*{a. Secret key rate}

\begin{figure}
\centering
\includegraphics[width=1\linewidth]{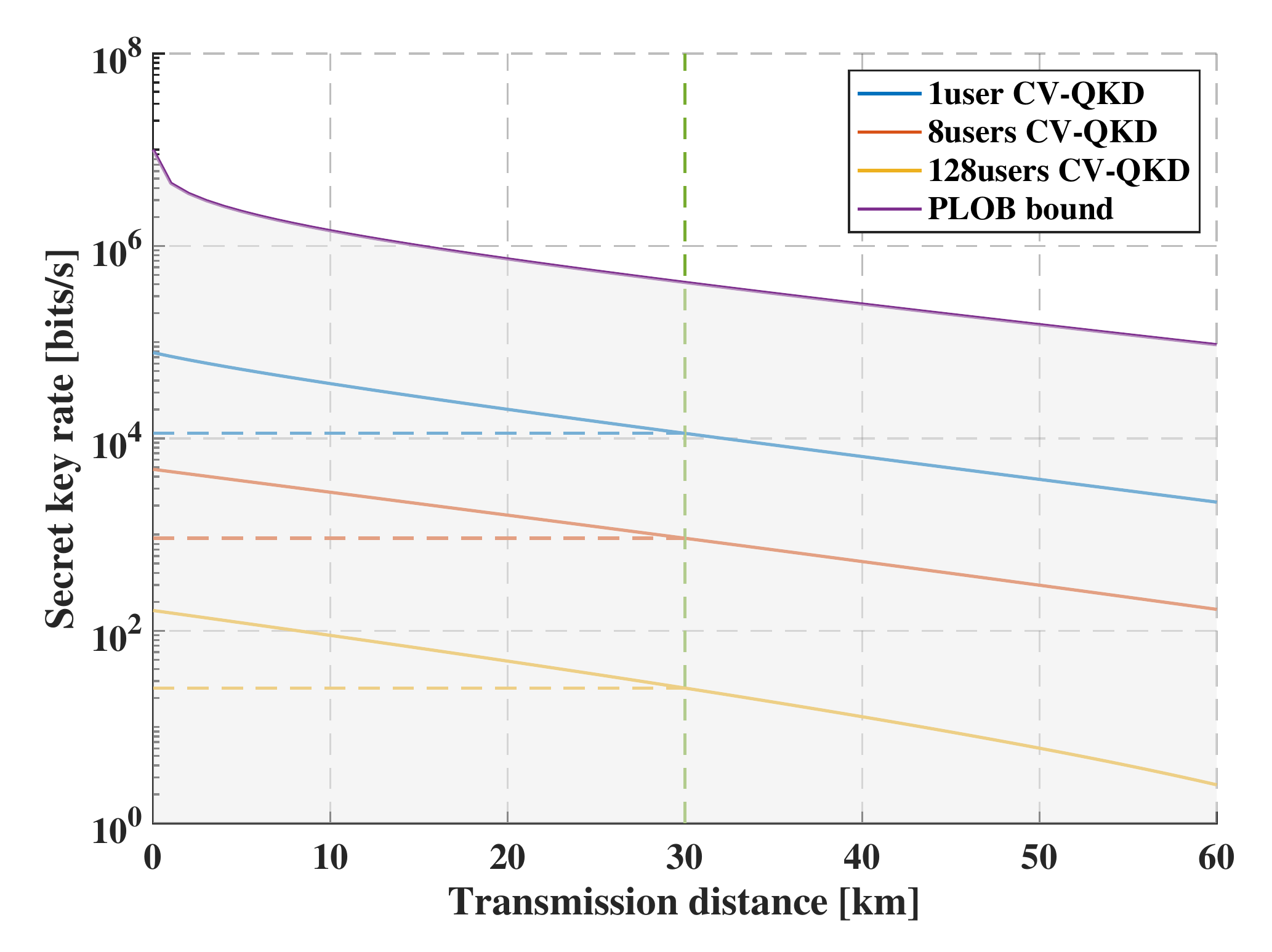}
\caption{The comparison diagram of secret key rate between this scheme of different network capacity and other classical schemes. The parameters set as $\eta = 0.42$, $v_{el}=0.18$, $\beta=0.97$, $V_{\rm{A}}=0.5 \;\rm{SNU\;(shot~noise~unit)}$, $R=1 \;\rm{MHz}$. It describes the change of secret key rate at different transmission distances in Pirandola-Laurenza-Ottaviani-Banchi (PLOB) bound and the round-trip multi-band continuous-variable quantum key distribution (CV-QKD) with different users, where the ordinate value corresponding to the dotted line is the secret key rate under the condition of $30\;\rm{km}$ achieved in theory.}
\label{fig8ex}
\end{figure}

\indent In the following, we evaluate the reachable secret key rate under the RM-QAN for discrete modulation coherent state (DMCS) CV-QKD \cite{leverrier2009unconditional}. The formula of the secret key rate for unit system repetition frequency is in the Appendix. For the practical CV-QKD system, the secret key rate $K_{s}$ can be calculated as
\begin{equation}
K_{s}=R K_{p},
\end{equation}
where $R$ is the repetition frequency of the CV-QKD system. The excess noise is the untrusted noise in the system. Through the physical noise analysis in Discussion, the excess noise of the RM-QAN can be described as
\begin{equation}
\label{eq5}
\varepsilon=\varepsilon_{\rm{RB}}+\varepsilon_{\rm{FC}}+\varepsilon_{\rm{OC}}+\varepsilon_{\rm{MO}}+\varepsilon_{\rm{AM}}+\varepsilon_{\rm{PH}}.
\end{equation}
In Eq. \ref{eq5}, $\varepsilon_{\rm{RB}}$ represents the noise introduced by Rayleigh backscattering. $\varepsilon_{\rm{FC}}$ denotes the noise introduced by the frequency crosstalk, which is caused by the photons leaking from other frequency bands. In addition, $\varepsilon_{\rm{OC}}$ represents the noise introduced by the imperfection of the optical circulator, especially concerning the isolation and directionality. $\varepsilon_{\rm{MO}}$ is the modulation noise caused by the uncertainty of modulation voltage. The change in the number of photons caused by the spontaneous radiation will be reflected in the amplitude of the laser, forming amplitude noise $\varepsilon_{\rm{AM}}$. Besides, the spontaneous radiation of the laser not only causes the change of intensity but also causes the random change in frequency of laser pulse signal, forming phase noise $\varepsilon_{\rm{PH}}$. Specific theoretical derivation values of each noise component can be found in Physical noise analysis in Discussion.

\indent Other parameters are quantum efficiency $\eta = 0.42$, electrical noise $v_{el}=0.18$, reconciliation efficiency $\beta=0.97$, modulation variance $V_{\rm{A}}=0.5 \;\rm{SNU}$ and repetition frequency $R=1 \;\rm{MHz}$. The N:1 splitter on the return path would introduce 1/N loss on each arm, thereby reducing the secret key rate of all user and the transmittance will be changed to $T=10^{-\alpha L / 10} / N_{S}$, where $N_{S}$ is the branch number of the N:1 splitter. The comparison diagram of the secret key rate between this scheme of different network capacity $N$ and other classical schemes is shown as FIG. \ref{fig8ex}.

\begin{figure*}
\centering
\subfigure[]{\label{fig9exa}\includegraphics[width=0.3\linewidth]{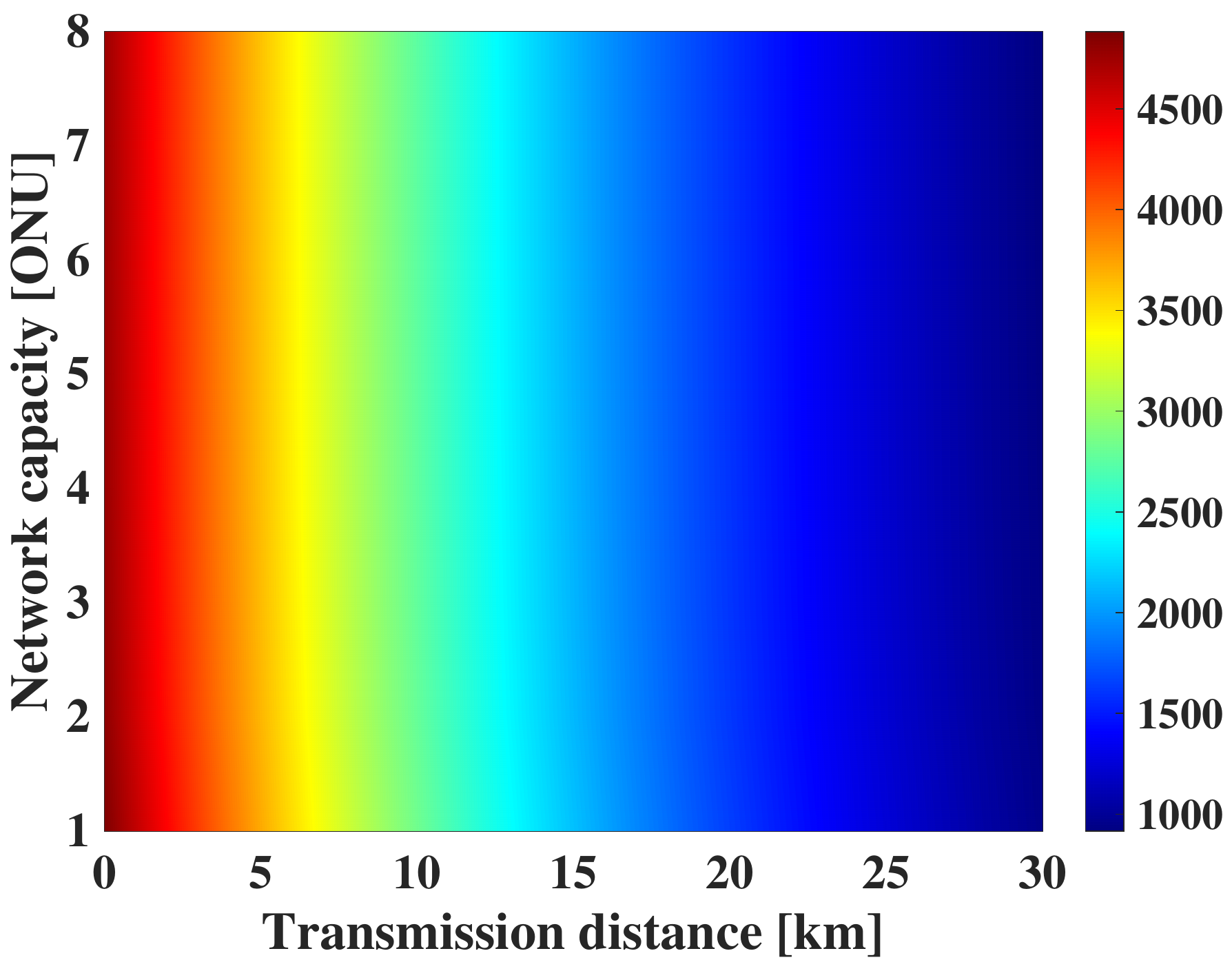}}
\hspace{3ex}
\subfigure[]{\label{fig9exb}\includegraphics[width=0.3\linewidth]{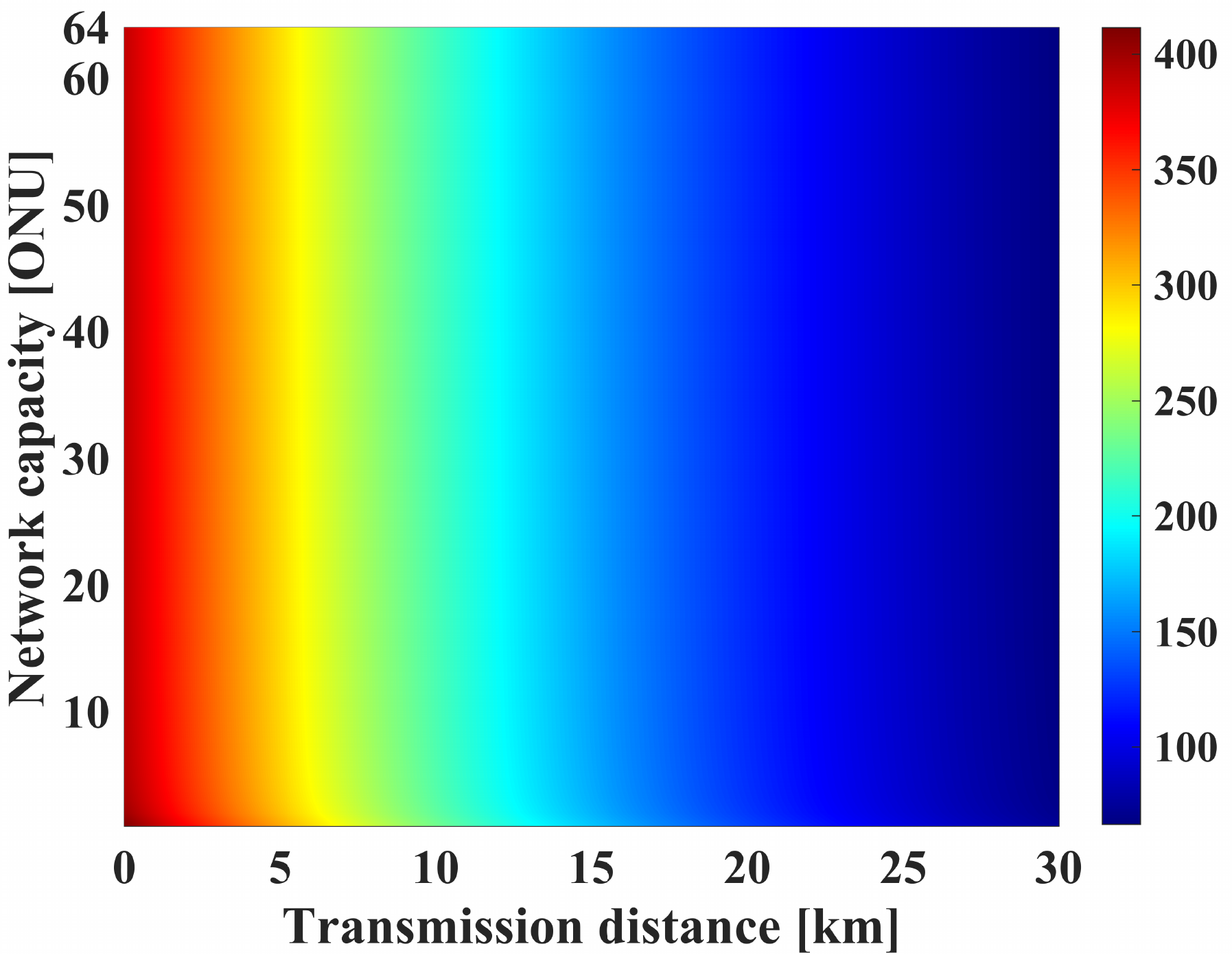}}
\hspace{3ex}
\subfigure[]{\label{fig9exc}\includegraphics[width=0.3\linewidth]{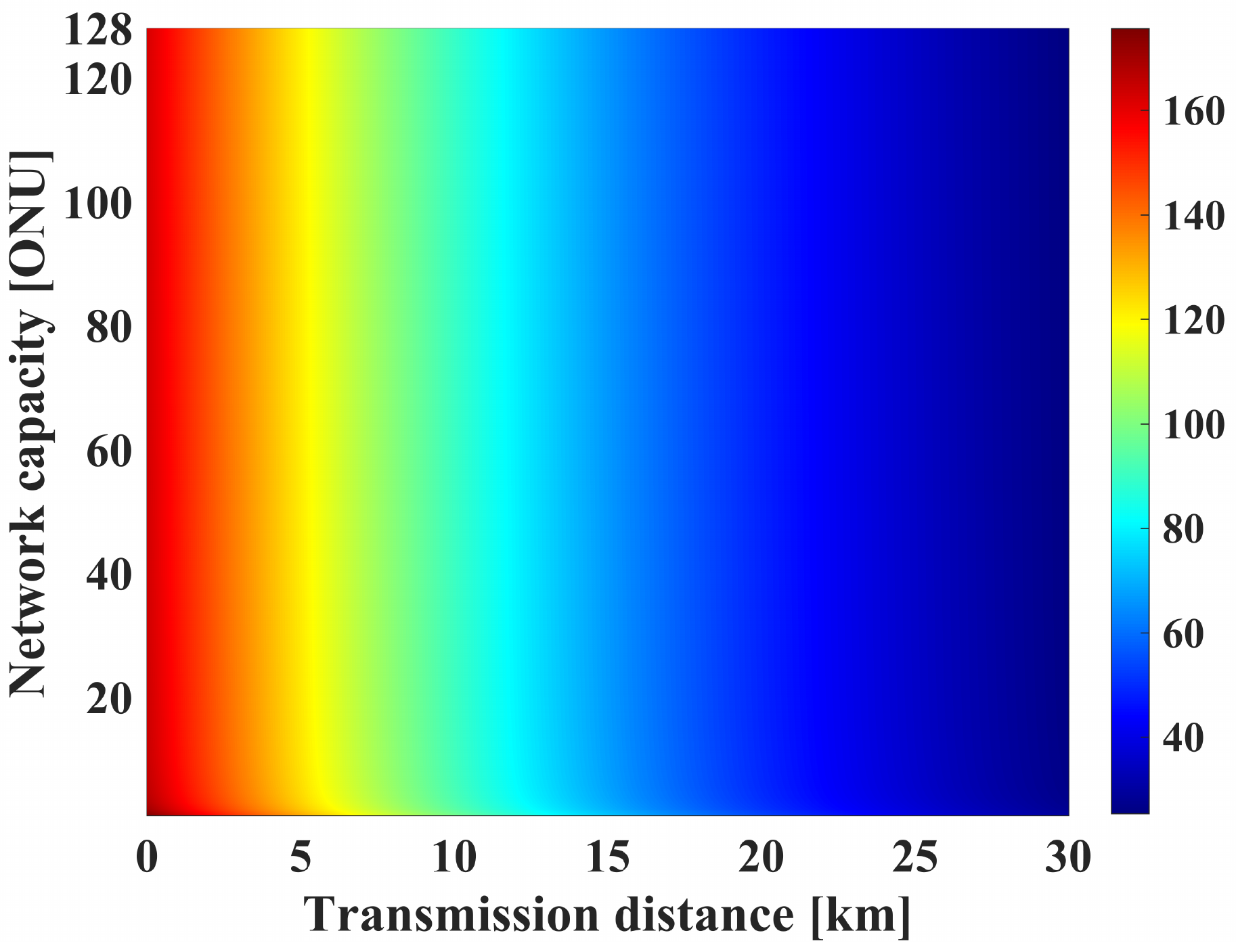}}
\caption{The relationship between network capacity, transmission distance, and secret key rate. The parameters set as $\eta = 0.42$, $v_{el}=0.18$, $\beta=0.97$, $V_{A}=0.5 \;\rm{SNU}$, $R=1 \;\rm{MHz}$. It describes the change of secret key rate at different transmission distances and different network capacities. The legend on the right shows the value of the secret key rate of each user. (a)The network capacity is 8. (b)The network capacity is 64. (c)The network capacity is 128.}
\label{fig9ex}
\end{figure*}

\indent In the above figure, Pirandola-Laurenza-Ottaviani-Banchi (PLOB) bound has the furthest transmission distance when the secret key rate is determined without a repeater and the highest secret key rate when the transmission distance is determined \cite{pirandola2017fundamental}. The curve of CV-QKD in different users is the relationship between the secret key rate and the distance corresponding to the different network capacities of RM-QAN. As can be seen from the figure, the secret key rate of all schemes decreases with the increase of transmission distance. Our scheme does not exceed the limit of PLOB under the same transmission distance. In addition, when the network capacity increases, the secret key rate decreases due to the gradual increase of the optical circulator noise and the frequency crosstalk noise. The secret key rate will decrease obviously with the increase of network capacity, since the assumed eavesdropper Eve can obtain $(N-1) /N$ signal, resulting in the increase of the maximum information that Eve can get. However, RM-QAN can still support the encoding of 128 users. It conforms to the concept of ``the last kilometer'' multi-access network.

\begin{figure*}[htb]
\centering
\includegraphics[width=1\linewidth]{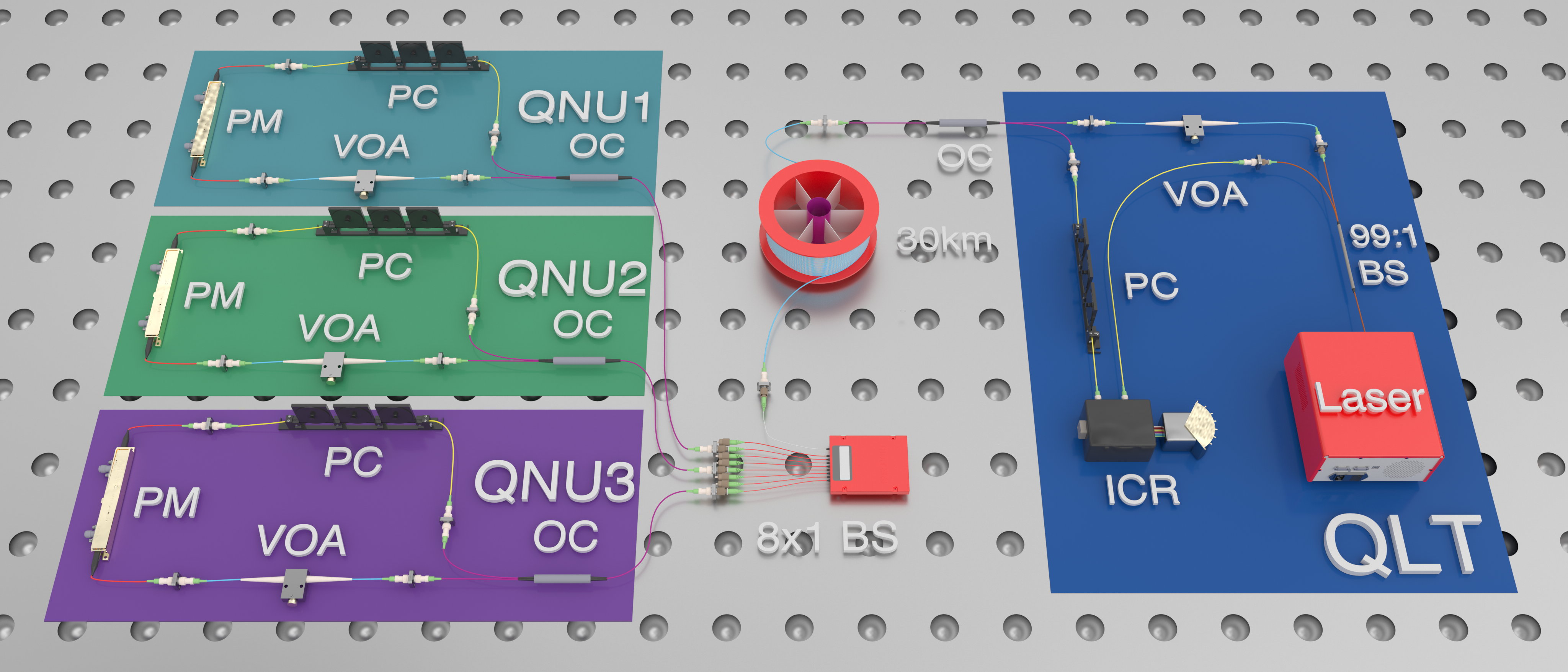}
\caption{The optical configuration of the round-trip multi-band scheme. First, the light transmitted by the laser of QLT is divided into two parts by a beam splitter (BS) with a 2:1 ratio. The light of higher power acts as the local oscillator (LO) of the system. The light of lower power passes through a variable optical attenuator (VOA) to the optical circulator. The light is transmitted from port1 to port2 of the optical circulator to the 30km optical fiber spool. Afterward, the continuous wave is transmitted to a BS through the optical fiber spool. After arriving at the QNU, the light is transmitted to a VOA through port2 to port3 of the optical circulator. The light then enters a phase modulator (PM) for signal modulation. Afterward, the signal light enters the polarization controller (PC). After passing through the PC, it enters through port1 of the optical circulator, exits through port2, and returns to BS. After passing through the optical fiber spool again, the signal light enters from port2 of the optical circulator at the QLT and comes out from port3. Then the signal light reaches the PC of QLT. Finally, the signal light and LO light enter the integrated coherent receiver (ICR).}
\label{fig10}
\end{figure*}

\subsubsection*{b. Network capacity}

\indent Network capacity is defined as how many users the network system can support. The network capacity is mainly affected by the branch number of the N:1 splitter. Meanwhile, the network capacity of RM-QAN is also affected by noise. The physical noise analysis in Discussion shows that only the optical circulator noise and the frequency crosstalk noise are related to the network capacity $N$. When the network capacity is small, frequency crosstalk noise is the main component of excess noise. However, when the network capacity is large, the frequency crosstalk noise tends to be constant, and the optical circulator noise is the main part of the noise. According to the previously calibrated parameters, the relationship between network capacity, transmission distance, and the secret key rate is shown in FIG. \ref{fig9ex}.

\indent In the above figure, the abscissa represents the transmission distance, the ordinate represents the network capacity, and the color legend on the right of the graph represents the value of the secret key rate of each user, which decreases gradually from red to blue. As can be seen from FIG. \ref{fig9ex}, when the network capacity is determined, the secret key rate decreases gradually with the increase of transmission distance. When the transmission distance is determined, the secret key rate decreases with the increase of network capacity. When users are added, the change of excess noise is small, and the secret key rate of each user remains almost constant. Certainly, if the detector bandwidth is not high enough, the number of users will be mainly limited by the detector bandwidth under the FDM scheme. If the number of users is larger, QLT requires higher output optical power to ensure that QNU achieves a certain output modulation variance, and we can use a high output power laser. It reflects that the round-trip multi-band quantum network scheme can accept a high network capacity without a performance penalty. In practical implementation, in order to achieve more accurate modulation variance adjustment, the light entering QNU and the total optical power will be larger, so the network capacity will decrease. If network capacity is determined, the secret key rate of each user will be stable and not affected when more users access into this network. In addition, our scheme only needs an optical circulator and a phase modulator to plug in a user, which has a low-cost requirement for new users to access and a small impact on the secret key rate of each existing user. In conclusion, this scheme has high practicability. 

\begin{figure*}
\centering
\subfigure[]{\label{fig11a}\includegraphics[width=0.3\linewidth]{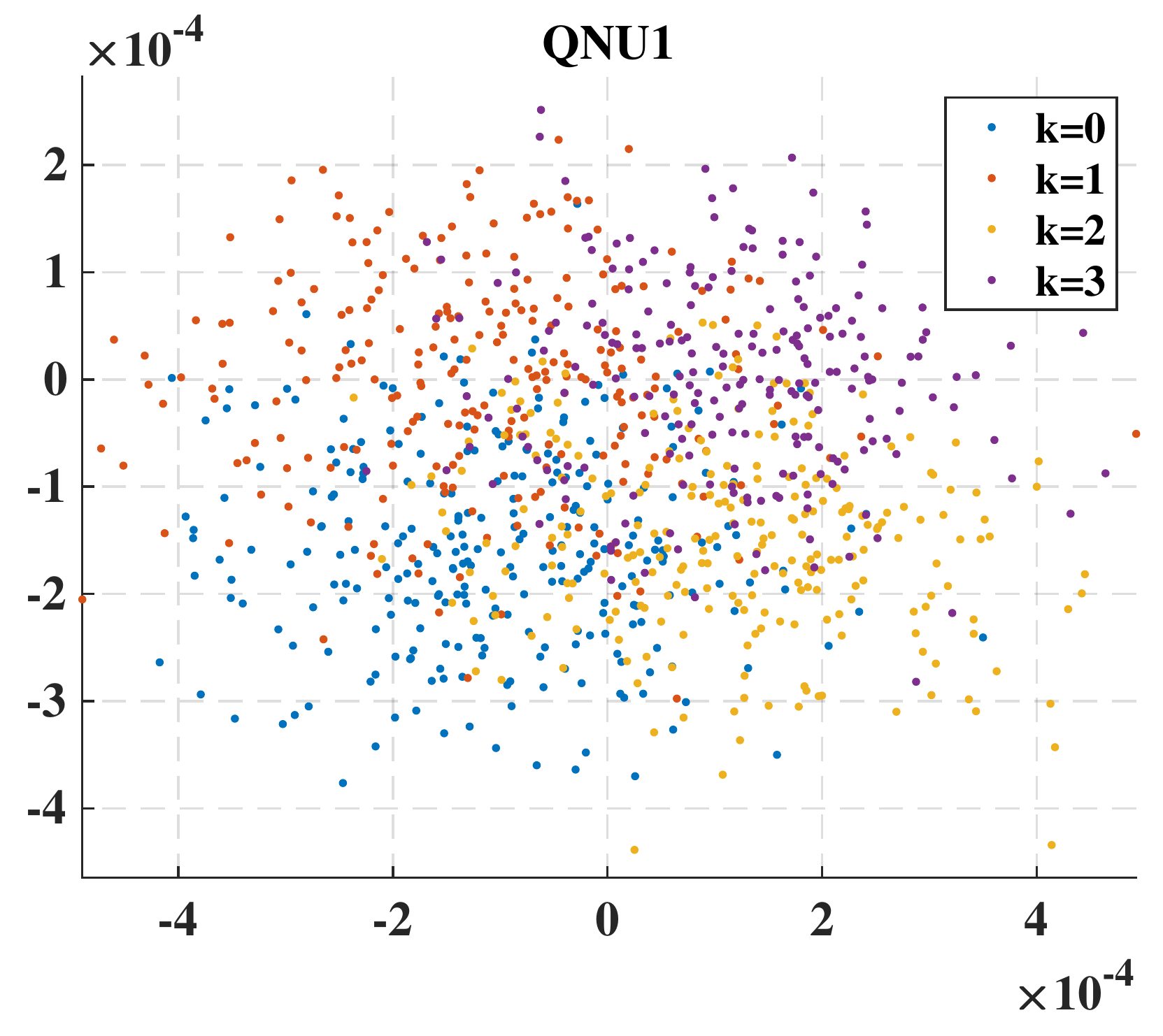}}
\hspace{3ex}
\subfigure[]{\label{fig11b}\includegraphics[width=0.3\linewidth]{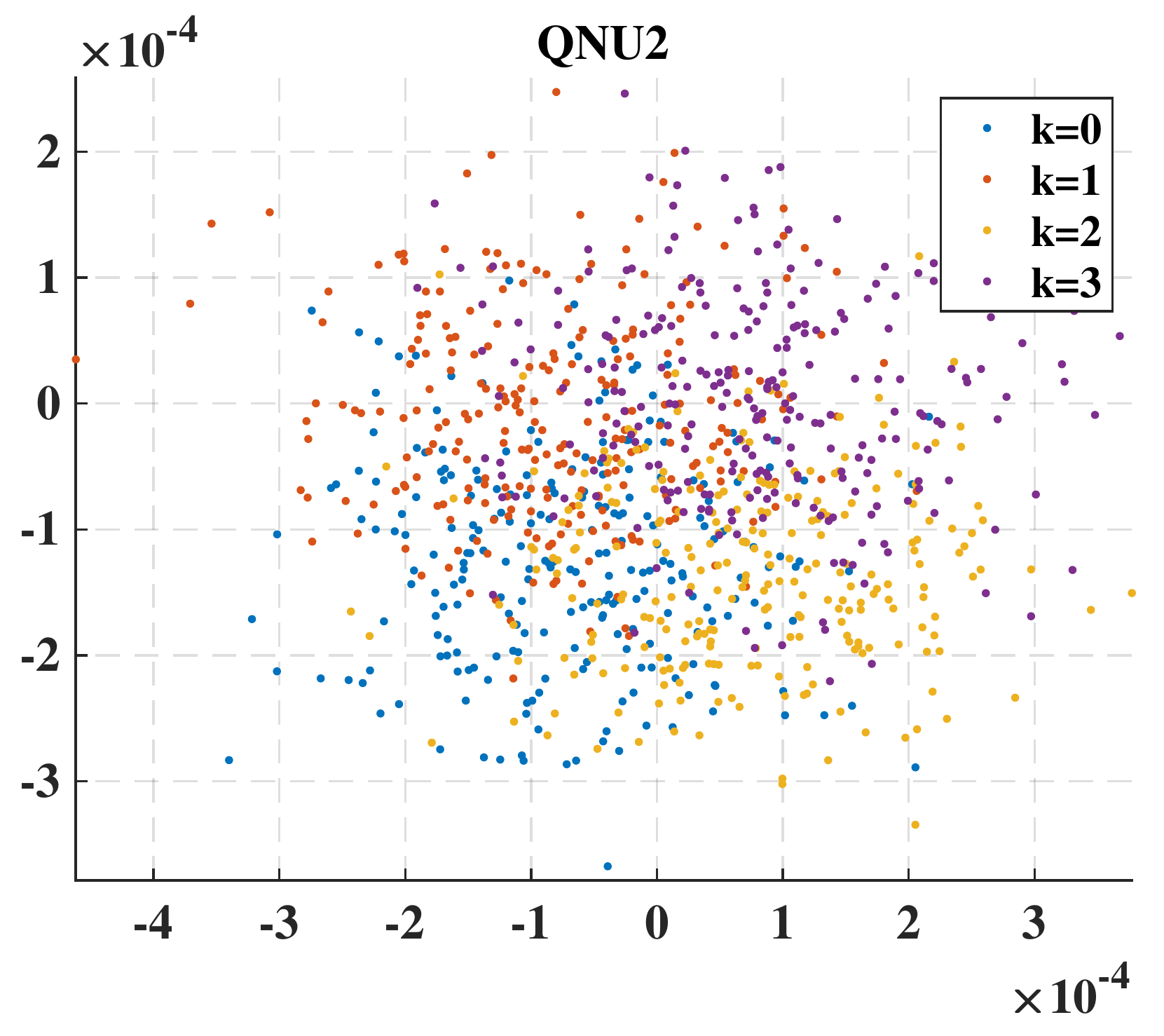}}
\hspace{3ex}
\subfigure[]{\label{fig11c}\includegraphics[width=0.3\linewidth]{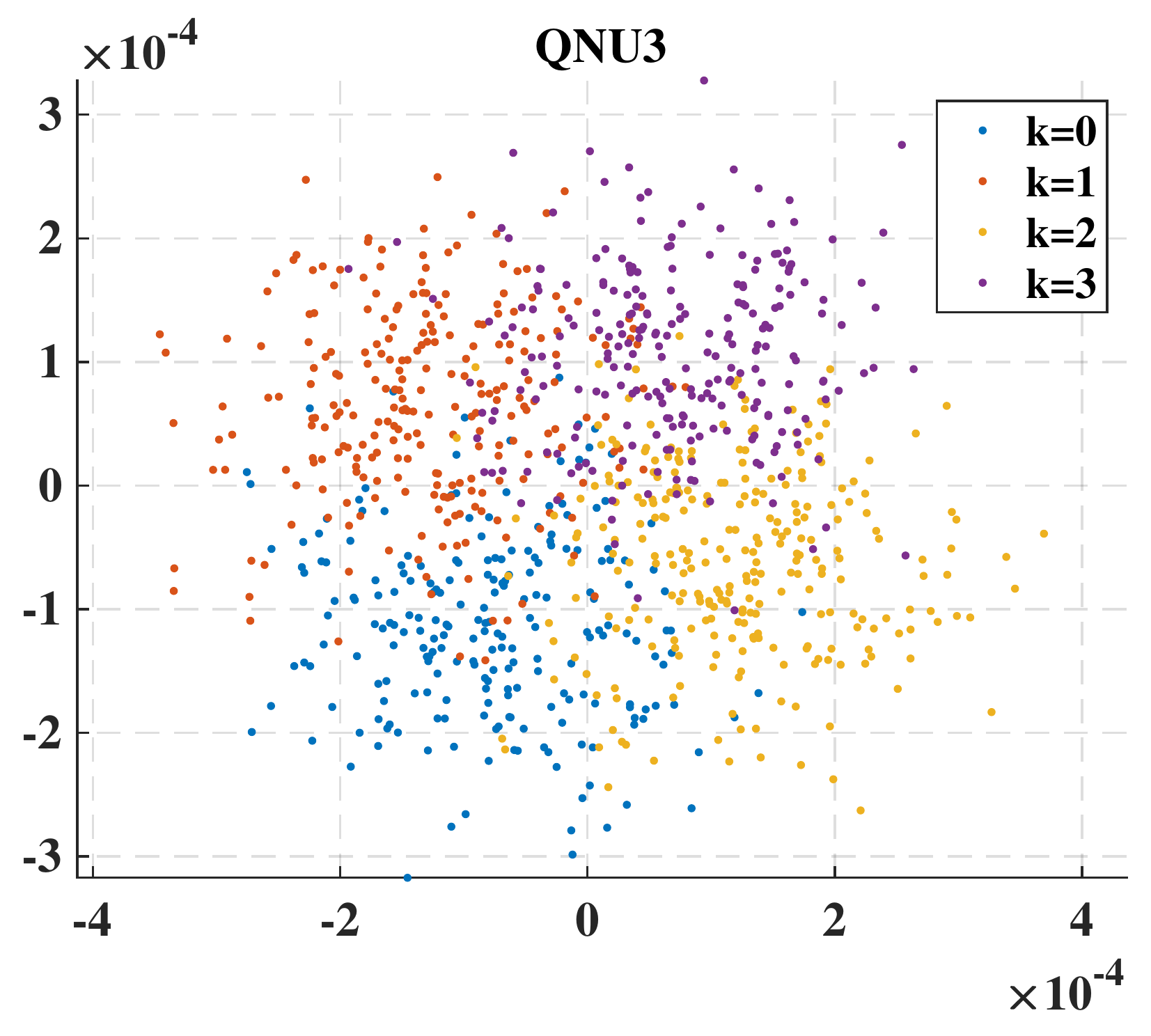}}
\caption{(a)The signal constellation of QNU1, with $ V_{A}=0.5587 \;\rm{SNU}$. (b)The signal constellation of QNU2, with $ V_{A}=0.5170 \;\rm{SNU}$. (c)The signal constellation of QNU3, with $ V_{A}=0.5641 \;\rm{SNU}$. The data in the figure were obtained through several experiments, where $k$ represents different states in discrete modulation coherent state (DMCS) CV-QKD. Different values of $k$ correspond to different colors.}
\label{fig11}
\end{figure*}

\begin{figure}[htb]
\centering
\includegraphics[width=1\linewidth]{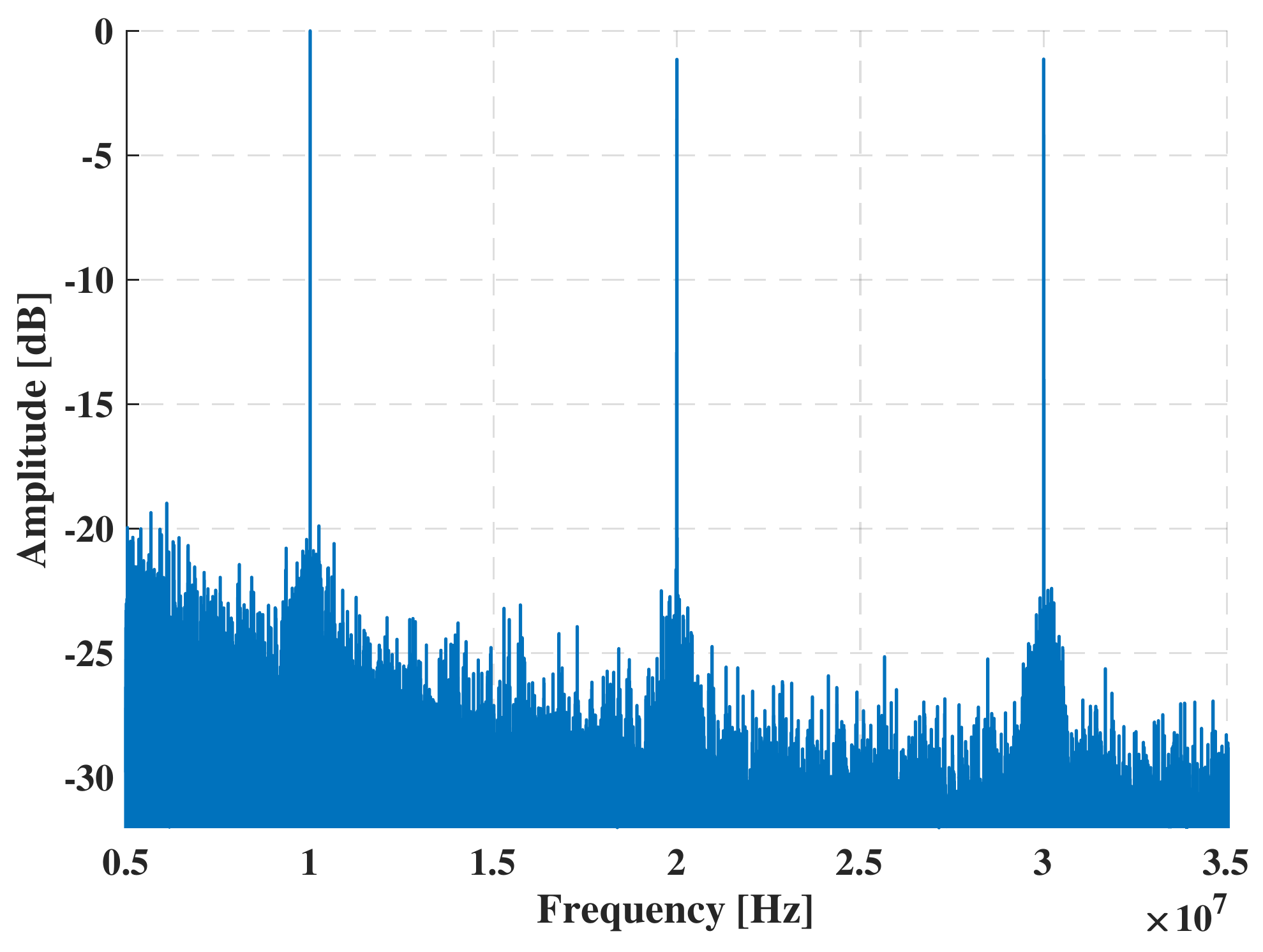}
\caption{Spectrum diagram of the signal obtained by QLT through the coherent receiver. It describes the change of amplitude at different frequencies.}
\label{fig12}
\end{figure}

\begin{figure}[htb]
\centering
\includegraphics[width=1\linewidth]{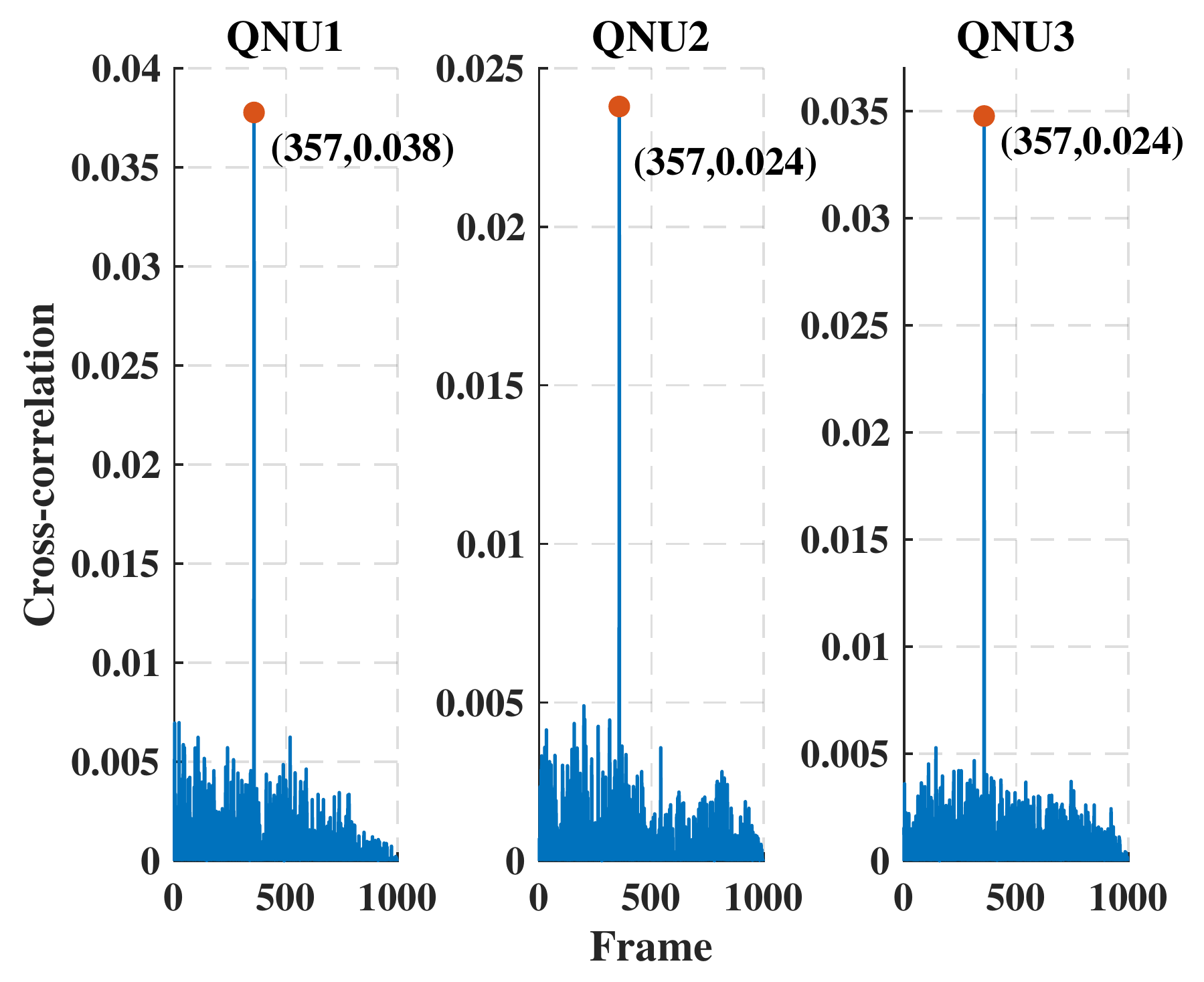}
\caption{The cross-correlation of the signal modulated by three QNUs and the signal received by QLT. It describes the change of the cross-correlation of three QNUs at different frames, where the red points represent the successful result.}
\label{fig13}
\end{figure}

\subsection*{Experiment verification}

\subsubsection*{a. Experimental set-up}

\indent The optical structure of RM-QAN used in the experiment is shown in FIG. \ref{fig10}. First, the light generated by the laser of QLT is divided into two parts by a beam splitter (BS) with a 2:1 ratio. The laser we use in the experiment has a very narrow linewidth, which is typical 100Hz. Therefore, the interval between the light before transmission and after transmission is less than the coherent length of the laser. On this basis, we realize homodyne detection, which means the beat frequency is 0. The light of higher power acts as the LO of the system. The light of lower power passes through a variable optical attenuator (VOA) to the optical circulator. The function of QLT's VOA is to weaken light into appropriate optical power. The light is transmitted from port1 to port2 of the optical circulator to the 30km optical fiber spool. Afterward, the continuous wave laser is transmitted to a BS through the optical fiber spool. Then the light is evenly divided into 8 QNUs by BS with a ratio of 8:1. Due to the limitation of modulated RF signal ports, we connected 3 QNUs in one experiment, and verified the experimental possibility of 8 users through multiple experiments. After arriving at the QNU, the light is transmitted to a VOA through port2 to port3 of the optical circulator. The function of QNU's VOA is to balance the optical power of each user. However, due to the manual adjustment of VOA, the optical power of each user cannot be completely identical, so there are differences in the SNR of each user in the experimental results. The light then enters a phase modulator (PM) for QKD modulation, which is achieved using an arbitrary waveform generator (AWG). Different QNUs modulate the information on different carrier frequencies, which can be distinguished on the spectrum diagram. QNU1, QNU2, and QNU3 in FIG. \ref{fig10} select 10MHZ, 20MHZ, and 30MHZ bands respectively. Afterward, the signal light enters the polarization controller (PC). The function of QNU’s PC is to adjust the polarization of each user, which can be eliminated after we use all polarization-maintaining fiber in practical implementation. After passing through the PC, it enters through port1 of the optical circulator, exits through port2, and returns to BS with an optical ratio of 8:1. At BS, the signal lights of the QNUs are converged into an optical fiber. After passing through the optical fiber spool again, the signal light enters from port2 of the optical circulator at the QLT and comes out from port3. Then the signal light reaches the PC of QLT for overall polarization adjustment. Finally, the signal light and LO light enter the integrated coherent receiver (ICR). After the detection, the signal is collected in an oscilloscope. According to the Nyquist sampling theorem, the sampling frequency must be higher than two times the highest signal frequency. In this case, there is no signal spectrum aliasing, and the signal can be completely recovered.

\begin{figure}
\centering
\includegraphics[width=1\linewidth]{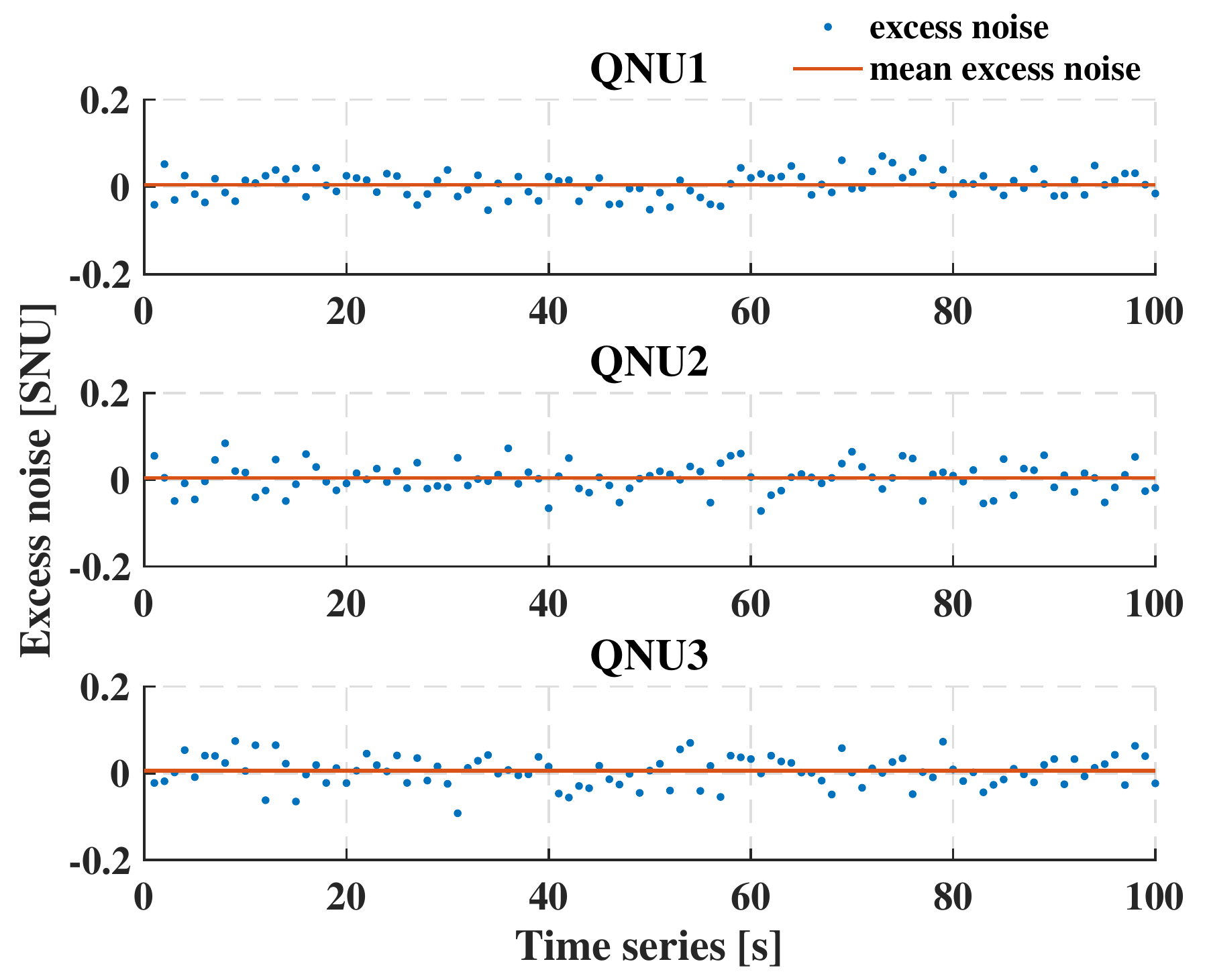}
\caption{The excess noise scatter diagram of three QNUs. It describes the change of the excess noise of three QNUs at different time, where the red line represents the mean value of excess noise of three QNUs. The mean excess noise of QNU1, QNU2 and QNU3 is $0.0054\;\rm{SNU}$, $0.0040\;\rm{SNU}$ and $0.0059\;\rm{SNU}$.}
\label{fig14}
\end{figure}

\subsubsection*{b. Signal processing}

\indent In order to separate the data of different users, FDM is adopted in the experiment. In the modulation of QNU, the modulated signal is the waveform generated by the baseband signal and corresponding carrier frequency. The signal can be represented by the Eq. \ref{ft}. Then the AWG loaded the signal $f(t)$ to the PM for phase modulation. There are two advantages to operating like this. First, the signal formed in this way has an extremely strong ability to resist phase shift and noise. Second, the signal must be an orthogonal four-state CV-QKD signal after the coherent demodulation. Their relative positions will not be changed, only by the overall rotation, which is convenient for signal recovery. After the coherent demodulation, the signal must be an orthogonal four-state CV-QKD signal. Their relative position will not be changed but will only be rotated as a whole.

\indent In the demodulation of QLT, bandpass filtering is firstly employed for different frequency bands. Afterward, QLT determines whether there is a signal on each registered frequency band. After that, the first phase shift recovery is carried out to recover the optical phase. Since the obtained signal is adopted spectrum shifting, QLT uses the coherent demodulation to restore the original signal of the baseband. It should be noted that the coherent demodulation of QLT needs to be used for $g_{a}(t)$ and $g_{b}(t)$. The formula of coherent demodulation can be expressed as Eq. \ref{gt}. QLT uses low-pass filtering of the baseband frequency to get the baseband signal. After downsampling, the second phase shift recovery is carried out to rotate the signal as a whole. Then, the signal is determined by cross-correlation to complete frame synchronization. In frame synchronization, it is necessary to pay attention to the effect of mean value on cross-correlation. After that, QLT will do the parameter estimation to evaluate the excess noise and further the information Eve can acquire, excluding this part in the post-processing. Besides, we will also monitor the input optical power at both QNU and QLT to close the potential practical security loophole.

\subsubsection*{c. Experimental results}

\indent The experimental results of RM-QAN are presented in the following. As can be seen from FIG. \ref{fig12}, the $10\;\rm{MHZ}$, $20\;\rm{MHZ}$, and $30\;\rm{MHZ}$ frequency bands modulated by three QNUs can be clearly seen in the mixed spectrum obtained by QLT. There is no spectrum aliasing in these bands. Since the modulation scheme is DMCS, the signal constellation presents four states, which correspond to the different values of $k$ in Eq. \ref{fu1}. Different values of $k$ correspond to different colors. As shown in constellation diagrams of three QNUs in FIG. \ref{fig11}, the signal variance gradually decreases to the variance of the shot noise with the reduction of SNR. It can also be seen from the constellation diagram that the four states become more and more indistinguishable with the decrease in SNR. However, the result of cross-correlation is still clear in FIG. \ref{fig13}, which can easily complete frame synchronization. Since AWG is a clock synchronized with the oscilloscope, each red point in FIG. \ref{fig13} should have the same horizontal coordinate, which means the frame synchronization position should be the same value. It was well verified in our experiments with excellent frame synchronization results. In addition, the excess noise of 100 frames of each QNU fluctuates around zero, as shown in FIG. \ref{fig14}. The excess noise mainly comes from frequency crosstalk noise, and the fluctuation comes from the deviation of data statistics. Their average is slightly more than zero. Inconsistent noise from different users results from statistical deviation in finite samples. Since we filter out noise from other users, the noise from different users will tend to be the same when the amount of data is large. In FIG. \ref{fig15}, the ordinate value corresponding to the dotted line is the achievable secret key rate under the condition of $30\;\rm{km}$ in our experiment, where the secret key rate of QNU1 is $825.82\;\rm{bits/s}$, the secret key rate of QNU2 is $674.46\;\rm{bits/s}$, and the secret key rate of QNU3 is $635.95\;\rm{bits/s}$. At present, due to the limitation of modulated RF signal ports, we have carried out experimental verification with eight users in multiple experiments. The feasibility of realizing more users can be fully inferred from the results, which can be realized by a BS with multiple interfaces.

\section*{Discussion}

\subsection*{Physical noise analysis}

\indent In RM-QAN, the total noise can be classified into trusted noises and untrusted noises according to the traditional QKD noise model. The trusted noises are the noises that can be calibrated by the receiver QLT, which cannot be controlled by Eve, and they constitute the electronic noise $v_{el}$ in the secret key rate bound calculation model. The untrusted noises, however, are caused by channel non-idealities or device non-idealities. These noises can not be calibrated accurately and are controllable by Eve. Therefore, these noises constitute the excess noise $\varepsilon$ in the secret key rate bound calculation model and indicate the extent of eavesdropping. In order to ensure the security of the QKD access network, we performed an analysis of actual physical noise. The untrusted noise introduced by our special scheme are Rayleigh backscattering noise $\varepsilon_{\rm{RB}}$, frequency crosstalk noise $\varepsilon_{\rm{FD}}$ and optical circulator noise $\varepsilon_{\rm{OC}}$. These are all untrusted noise. At the same time, we also briefly analyze the classical untrusted noise and the classical trusted noise in general CV-QKD. In this way, the system performance can be analyzed by this updated physical noise model.

\begin{figure}
\centering
\includegraphics[width=1\linewidth]{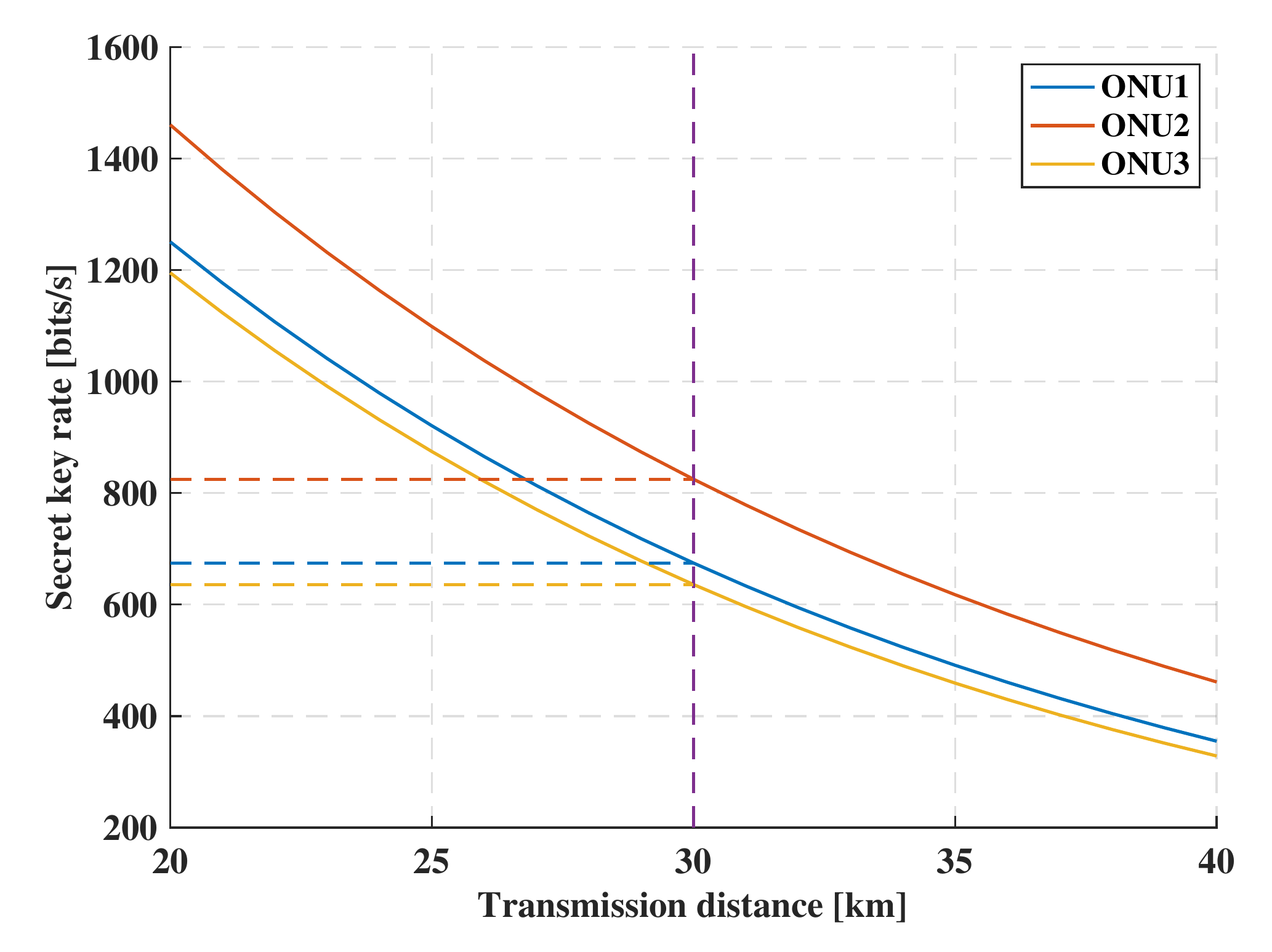}
\caption{The secret key rate curves of the three QNUs in the experiment. It describes the change of the secret key rate of three QNUs at different transmission distances, where the ordinate value corresponding to the dotted line is the secret key rate under the condition of $30\;\rm{km}$ achieved in our experiment. The secret key rate of QNU1 is $825.82\;\rm{bits/s}$, the secret key rate of QNU2 is $674.46\;\rm{bits/s}$, and the secret key rate of QNU3 is $635.95\;\rm{bits/s}$.}
\label{fig15ex}
\end{figure}

\subsubsection*{a. Rayleigh backscattering noise}

\indent In our scheme, light is transmitted from QLT by a round-trip structure, modulated by QNU, and received by QLT. This round-trip structure will have a scattering effect, resulting in noise. The scattering noise is mainly divided into Rayleigh backscattering noise and Raman scattering noise. Raman scattering noise mainly affects the wavelength division multiplexing (WDM) system \cite{fujiwara2006impact}. However, our FDM access network uses a single laser, which makes the wavelength unique. Moreover, theoretical analysis shows that the influence of Raman scattering on CV-QKD can be ignored because the LO light acts as a filter. In conclusion, only Rayleigh backscattering noise needs to be considered \cite{subacius2005backscattering}.

\indent Rayleigh backscattering noise $\varepsilon_{\rm{RB}}$ is caused by the interference of noise photons in the same spectral segment in the coherent detection of QLT, which can neither be filtered out nor attenuated. Rayleigh backscattering can occur anywhere in the fiber, and it cannot be monitored by a timing detector. Therefore, Rayleigh backscattering noise is something that needs to be considered in our scheme. The number of scattered photons produced by the Rayleigh backscattering effect is \cite{subacius2005backscattering}
\begin{equation}
	\left\langle\widehat{N}_{\mathrm{RB}}\right\rangle=(1-T) 10^{\beta / 10}\left\langle\widehat{N}_{\mathrm{QNU}}\right\rangle R,
\end{equation}
where $R$ is the repetition frequency of the system, and $\eta=10^{-L_{\mathrm{QNU}} / 10}$. Since all QNUs are in parallel, $L_{\mathrm{QNU}}$ is the loss inside one QNU round-trip. The transmittance is $T=10^{-\alpha L / 10}$, where $L$ represents the length of the optical fiber, $\alpha$ denotes the attenuation coefficient. $\beta$ represents the Rayleigh backscattering coefficient. $\left\langle\widehat{N}_{\mathrm{QNU}}\right\rangle=0.5V_{\mathrm{A}}$ is the number of photons returning from QNU, where $V_{\mathrm{A}}$ is the modulation variance. $\tau$ is the electric integration time of the heterodyne detector, namely the gate pulse time, then Rayleigh backscattering noise is
\begin{equation}
\begin{aligned}
	\varepsilon_{\rm{RB}}&=\frac{2\left\langle\widehat{N}_{\mathrm{RB}}\right\rangle \tau}{\eta T}\\&=\frac{2\left(1-10^{-\alpha L / 10}\right) 10^{\beta / 10}\left\langle\widehat{N}_{\mathrm{QNU}}\right\rangle R \tau}{\eta 10^{-\alpha L / 10}}\\&=\frac{\left(1-10^{-\alpha L / 10}\right) 10^{\beta / 10} V_{\mathrm{A}} R \tau}{10^{-L_{\mathrm{QNU}} / 10} 10^{-\alpha L / 10}}.
\end{aligned}
\end{equation}

\begin{figure}
\centering
\includegraphics[width=1\linewidth]{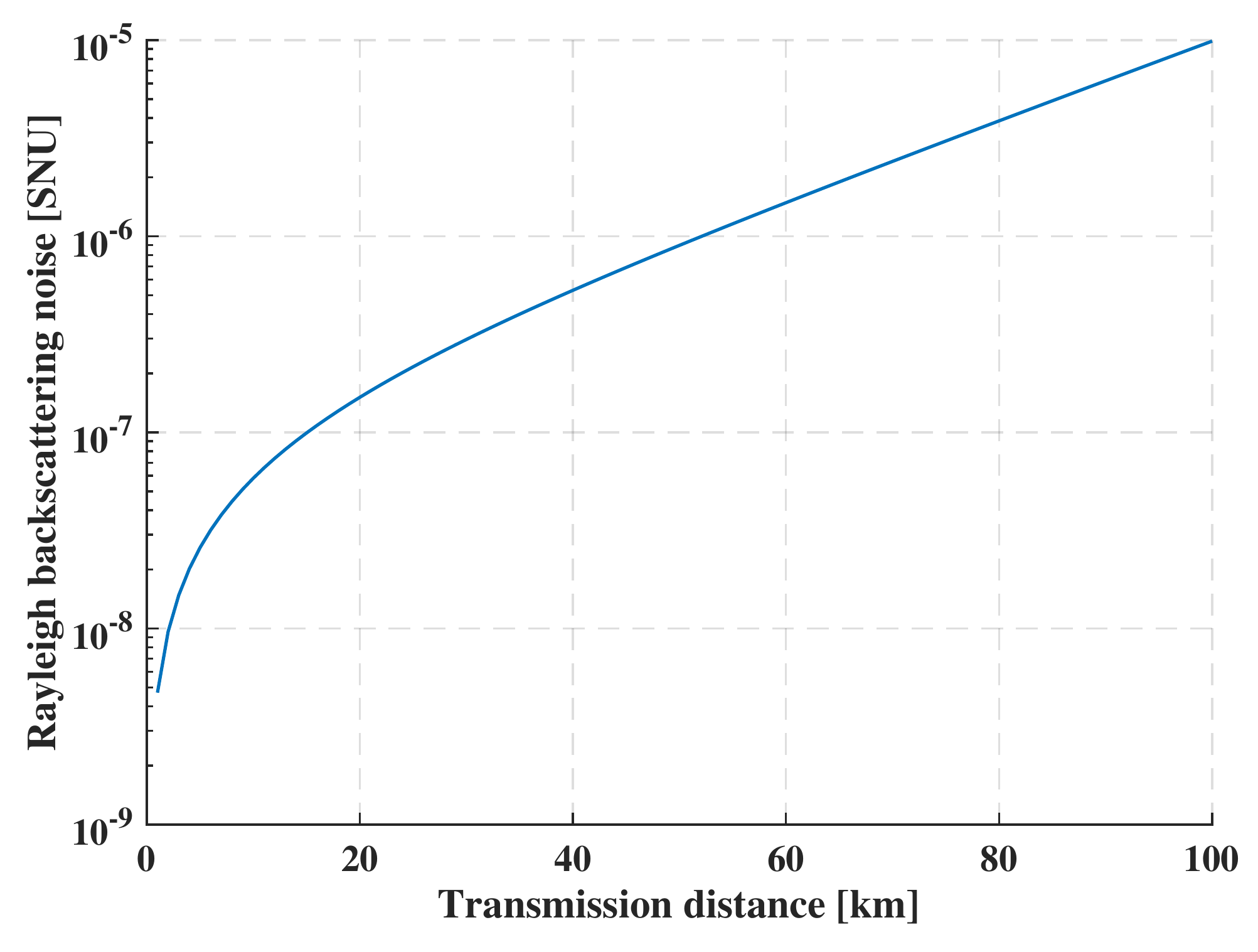}
\caption{The relationship between Rayleigh backscattering noise and transmission distance. The parameters set as $\beta=-40\;\rm{dB}$, $\alpha=0.2\;\rm{dB}/km$, $V_{\mathrm{A}}=0.5\;\mathrm{SNU}$, $R=1\;\mathrm{MHz}$, $\tau=1\;\mathrm{ns}$, $L_{\mathrm{QNU}}=3\;\mathrm{dB}$. It describes the change of Rayleigh backscattering noise at different transmission distances.}
\label{fig3}
\end{figure}

\indent As shown in FIG. \ref{fig3}, this image describes the relationship between Rayleigh backscattering noise $\varepsilon_{\rm{RB}}$ and transmission distance $L$. As can be seen from the figure, Rayleigh backscattering noise increases gradually with the increase of transmission distance, but its order of magnitude is always within an acceptable range.

\subsubsection*{b. Frequency crosstalk noise}

\indent Because our access network scheme uses FDM to distinguish users, it inevitably produces photon crosstalk between different frequency bands. When filtering the frequency band of a single user, the photons from other frequency bands leak in and produce noise. This noise is called frequency crosstalk noise $\varepsilon_{\rm{FC}}$.

\indent Before considering the frequency crosstalk noise of all other users to a single user, we can simply consider the noise between any two bands $\varepsilon_{\rm{F}}$. Firstly, frequency interval $\Delta f$ is an important parameter to describe noise between frequency bands $\varepsilon_{\rm{F}}$. In addition, the intensity of signal light also affects the interband noise $\varepsilon_{\rm{F}}$, which can be described by modulation variance $V_{\mathrm{A}}$. As the FDM scheme needs to use a bandpass filter to separate signals in different bands, the parameters of the filter also affect the noise between bands $\varepsilon_{\rm{F}}$. The influence of interband noise $\varepsilon_{\rm{F}}$ is also different for different filter types. The classical Butterworth filter, which is also the filter used in our experiment, will be analyzed in the following. Butterworth filter is mainly affected by the passband range and stopband range of the bandpass filter. It should be noted that other parameters of the filter, such as sampling rate, maximum passband attenuation, and maximum stopband attenuation, can be taken as reasonable values. Within reasonable limits, these parameters have little influence on filtering results. In conclusion, interband noise $\varepsilon_{\rm{F}}$ is mainly affected by frequency interval $\Delta f$, modulation variance $V_{\mathrm{A}}$, passband range, and stopband range.

\begin{figure}
\centering
\includegraphics[width=1\linewidth]{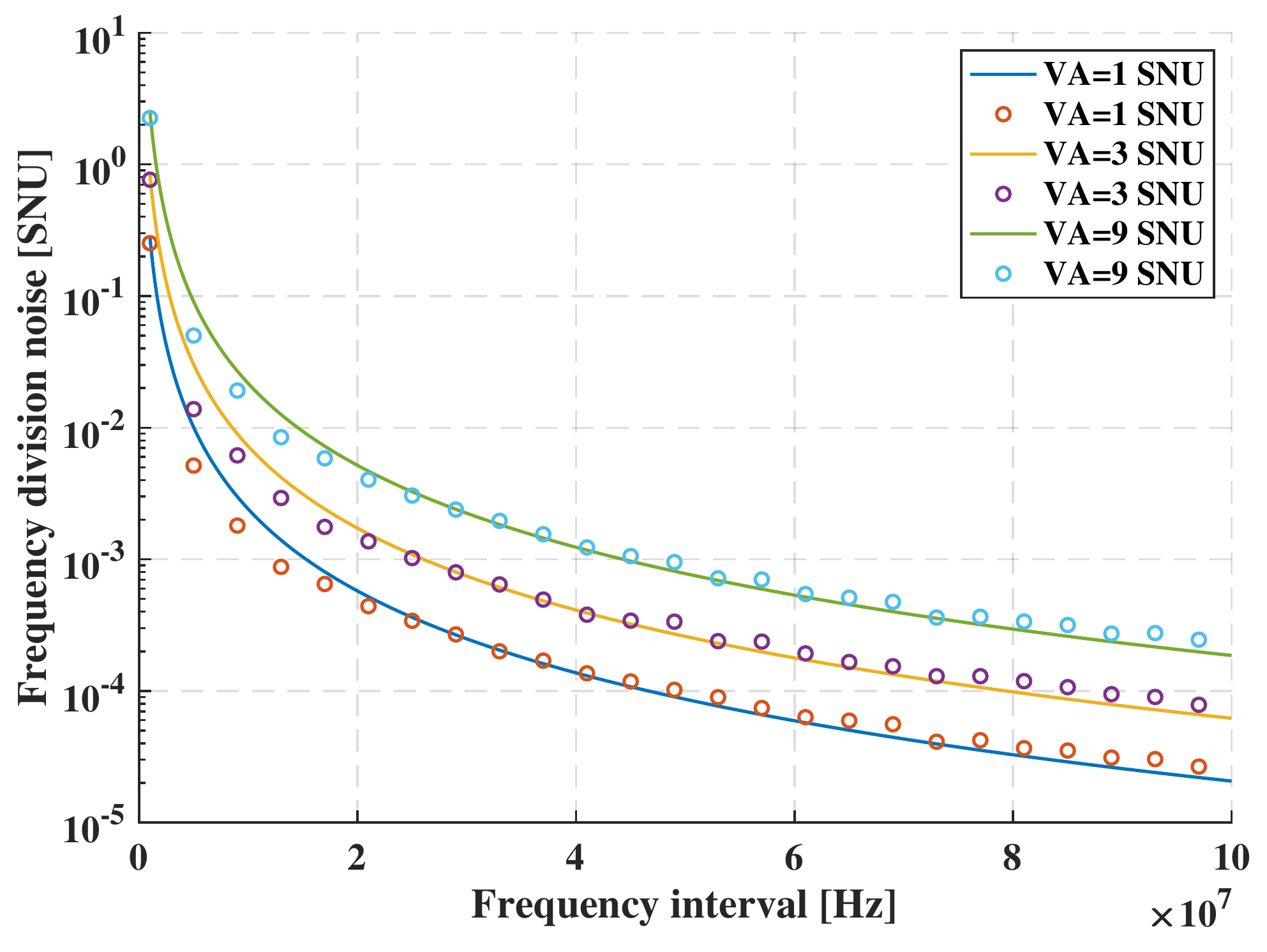}
\caption{The relation between frequency interval and frequency division noise under different modulation variances. The parameters set as $a=27.27, b=-2.066$. Hollow points are obtained by Monte Carlo method and curves are obtained by nonlinear fitting. It describes the change of frequency division noise at different frequency intervals and different modulation variances ($V_{\mathrm{A}}$).}
\label{fig4}
\end{figure}

\indent For the passband range and stopband range, which are difficult to be described by formulas, we use experimental methods to determine their better values. Of course, it is also possible to determine a better adaptive dynamic filter through machine learning and other algorithms to achieve a better filtering effect, but this is not within the scope of this paper. After hundreds of experiments, we got a reasonable passband and stopband range. Since signals mainly exist in the first main lobe, we take the passband boundary as the first main lobe of the signal and the stopband boundary as the second main lobe of the signal. This filtering result may not be the best, but it has met our experimental requirements.

\indent Since the filtering process is solving linear differential equations with constant coefficients, we use the Monte Carlo method to obtain the noise between frequency bands $\varepsilon_{\rm{F}}$. After several simulations, we obtained the relation between frequency interval $\Delta f$ and frequency division noise $\varepsilon_{\rm{F}}$ under different modulation variances $V_{\mathrm{A}}$, as shown in FIG. \ref{fig4}.

\begin{figure}
\centering
\includegraphics[width=1\linewidth]{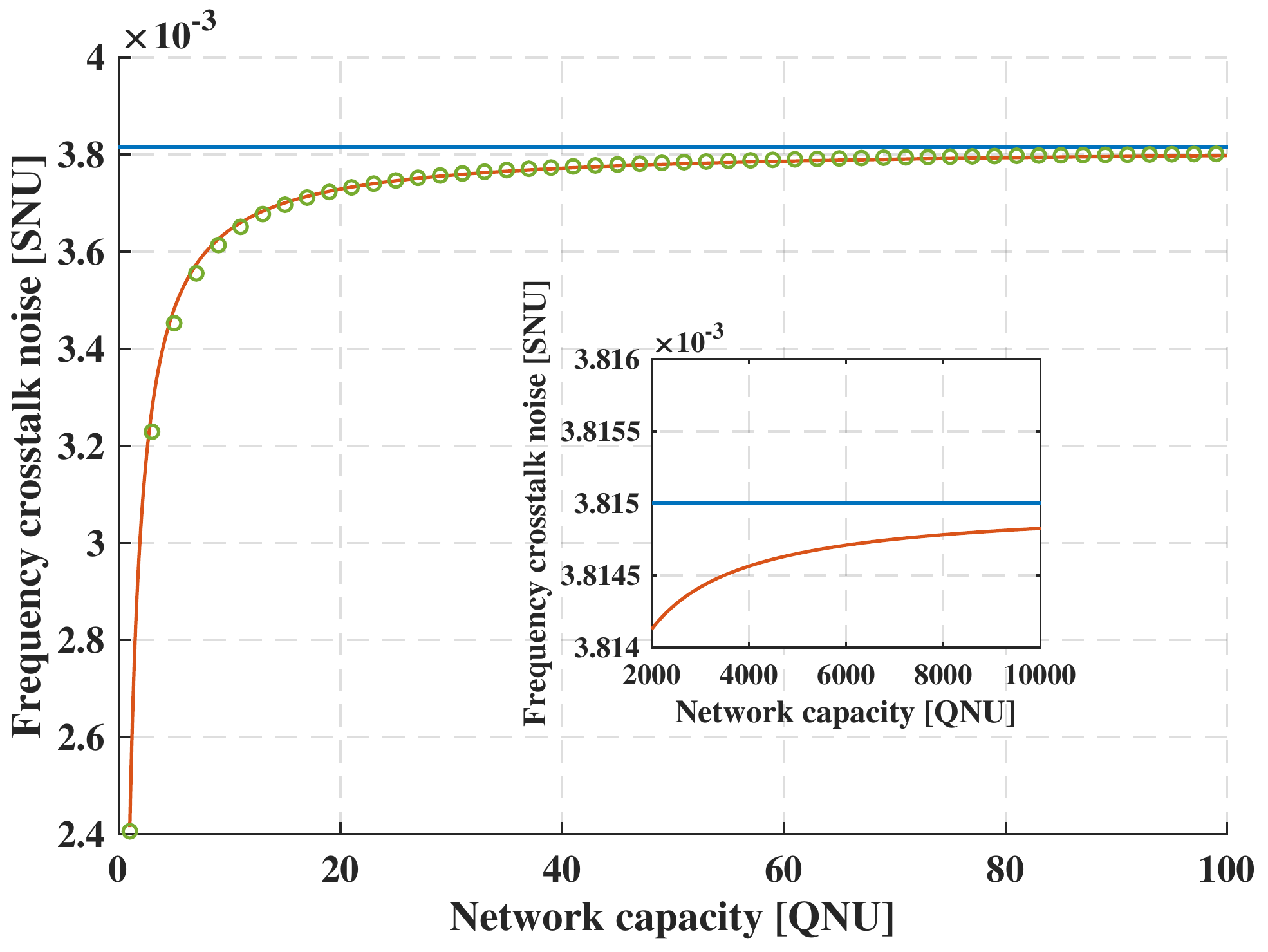}
\caption{The relation between network capacity and frequency crosstalk noise. The parameters set as $c=3.815 \times 10^{-3}, d=-0.4576$. Hollow points are obtained by the Monte Carlo method, and curves are obtained by nonlinear fitting. The small graph shows what happens when the network capacity is large. It describes the change of frequency crosstalk noise at different network capacities.}
\label{fig5}
\end{figure}

\indent The dots in the figure represent values obtained by the Monte Carlo method, and the curves are obtained by the nonlinear fitting method. It can be seen from the figure that the noise between frequency bands $\varepsilon_{\rm{F}}$ is proportional to the modulation variance $V_{\mathrm{A}}$ and decreases with the increase of frequency interval $\Delta f$. It is worth mentioning that the $V_{\mathrm{A}}$ is rounded to better represent it in FIG. \ref{fig4}. The value of $V_{\mathrm{A}}$ does not affect the generality of the resulting formula. Thus, interband noise $\varepsilon_{\rm{F}}$ can be expressed as
\begin{equation}
	\varepsilon_{\rm{F}}=V_{\mathrm{A}}\left(e^{a}+\Delta f^{b}\right),
\end{equation}
where $a=27.27, b=-2.066$. Given the noise between any two frequency bands $\varepsilon_{\rm{F}}$, we can calculate the frequency crosstalk noise $\varepsilon_{\rm{FC}}$ of a single user, which is affected by all the other users. In our experiment, we found that the noise $\varepsilon_{\rm{F}}$ when there are users on both sides at the same distance is the same as the noise $\varepsilon_{\rm{F}}$ when there are users on only one side. So frequency crosstalk noise $\varepsilon_{\rm{FC}}$ is only relevant to the presence or absence of the user at a certain distance. Thus, we get the influence of the number of users, namely network capacity $N$, on noise $\varepsilon_{\rm{FC}}$ when $V_{\mathrm{A}}=1\;\rm{SNU}$, as shown in FIG. \ref{fig5}.

\begin{figure*}[htb]
\centering
\includegraphics[width=1\linewidth]{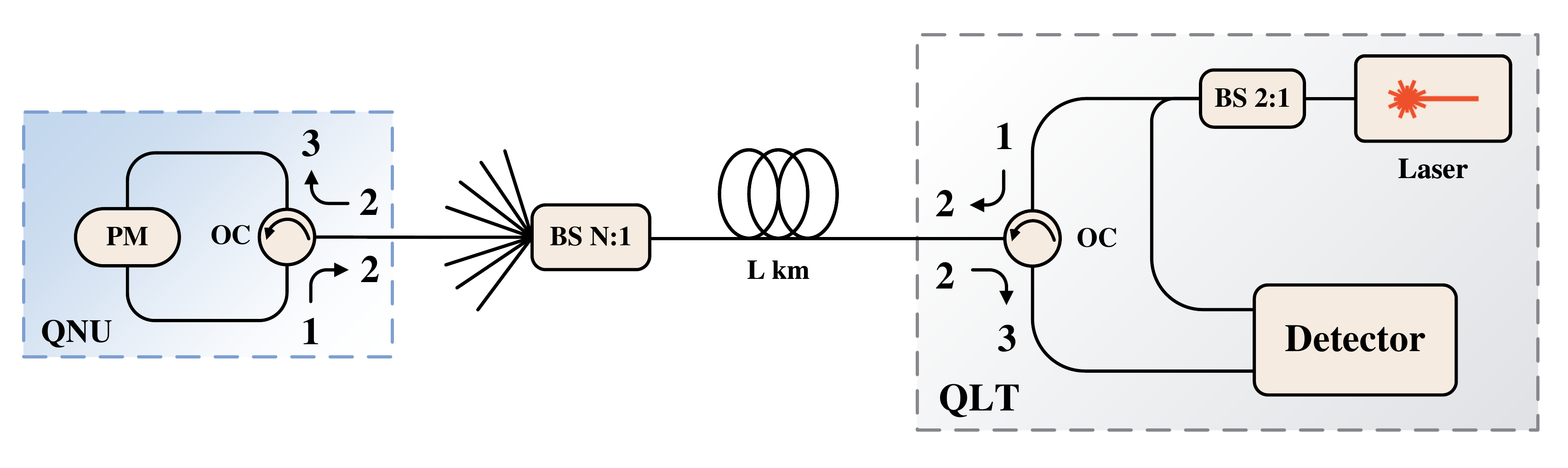}
\caption{Schematic diagram of optical circulator noise. The optical circulator of QLT receives the light from the laser at port1 and transmits it to QNU at port2. The signal light returned by QNU is then received at port2 and output to the detector at port3. The optical circulator of QNU receives the light from the QLT at port2, then transmits it to the PM at port3.}
\label{fig6}
\end{figure*}

\indent The dots in the figure represent discrete values of noise $\varepsilon_{\rm{FC}}$ under different network capacities, and the curves represent continuous values derived from the nonlinear fitting method. As can be seen from the figure, the frequency crosstalk noise $\varepsilon_{\rm{FC}}$ influenced by all other users increases gradually with the increase of network capacity, but the increase rate decreases gradually. It can be described as
\begin{equation}
	\varepsilon_{\rm{FC}}=V_{\mathrm{A}}ce^{d / N},
\end{equation}
where $c=3.815 \times 10^{-3}, d=-0.4576$. Thus, we can get that the noise $\varepsilon_{\rm{FC}}$ will gradually approach the value of $c$ with the increase of network capacity $N$, as shown in the small figure of FIG. \ref{fig5}. When the network capacity is $N=10000, \varepsilon_{\rm{FC}}$ is extremely close to $3.815 \times 10^{-3}$. When the network capacity tends to be infinite, $\varepsilon_{\rm{FC}}=3.815 \times 10^{-3}$.

\subsubsection*{c. Optical circulator noise}

\begin{figure*}
\centering
\subfigure[]{\label{fig7a}\includegraphics[width=0.47\linewidth]{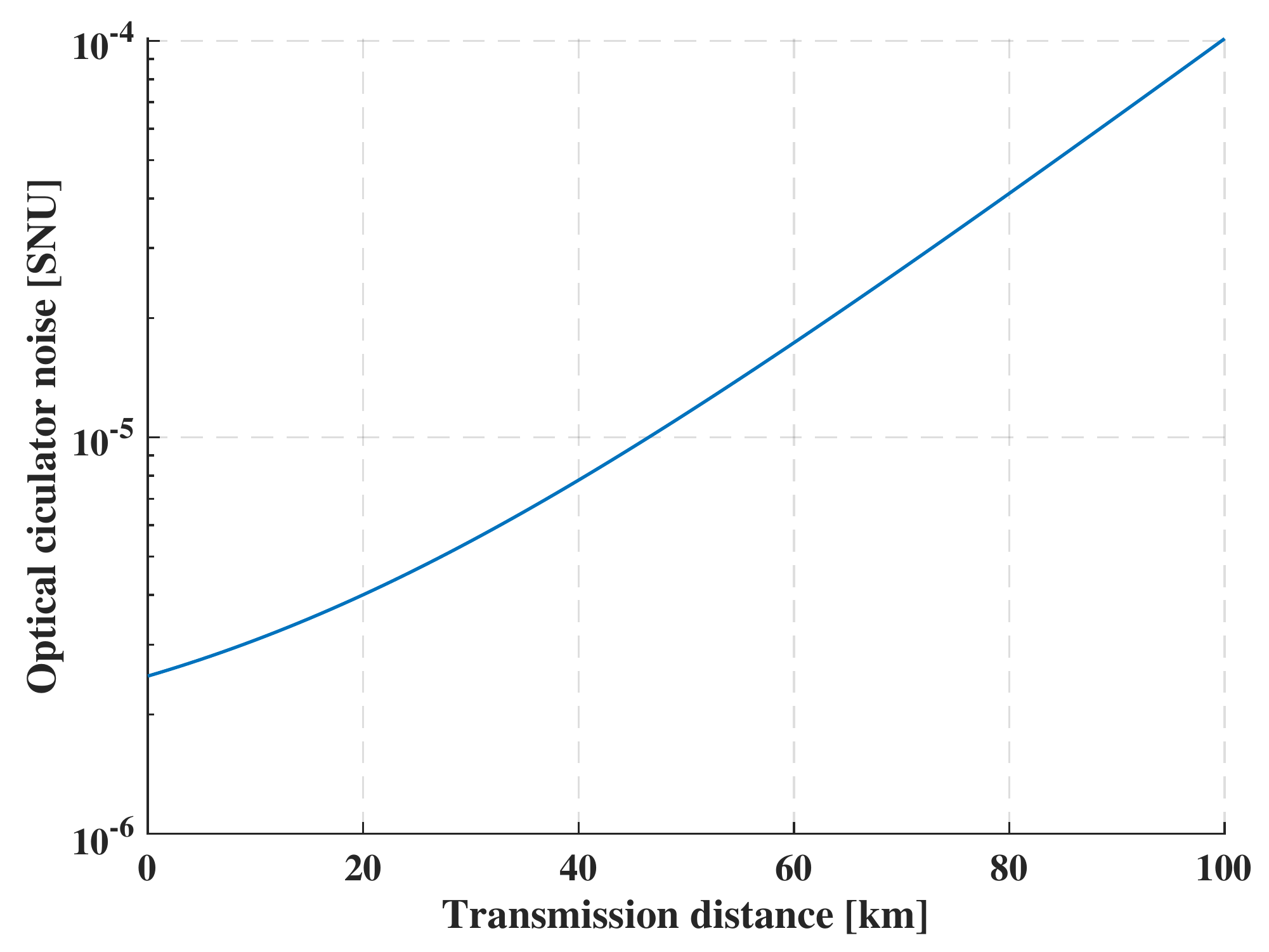}}
\hspace{3ex}
\subfigure[]{\label{fig7b}\includegraphics[width=0.47\linewidth]{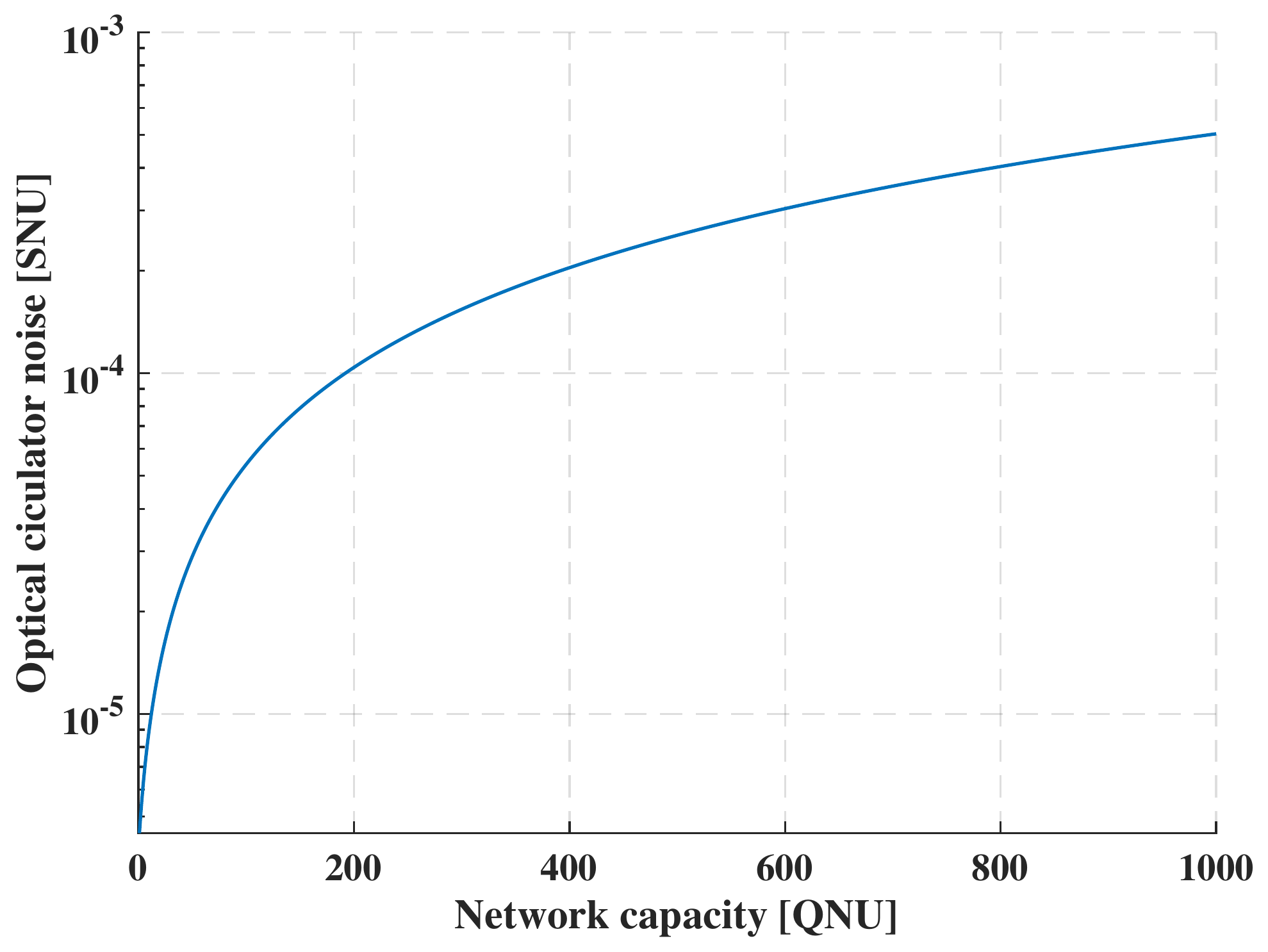}}
\caption{(a)The relation between optical circulator noise and transmission distance. The image describes the change of optical circulator noise at different transmission distances (b)The relation between optical circulator noise and network capacity. It describes the change of optical circulator noise at different network capacities. The parameters set as $D=60\;\rm{dB}$, $\alpha=0.2\;\rm{dB}/km$, $V_{\mathrm{A}}=0.5\;\mathrm{SNU}$, $L_{\mathrm{QNU}}=3\;\mathrm{dB}$.}
\label{fig7}
\centering
\end{figure*}

\indent The round-trip quantum access network introduces an optical circulator to complete the optical round-trip scheme. However, the optical circulator is not found in the classical CV-QKD system. Therefore, we analyze the noise effect of the optical circulator on the system.

\indent An optical circulator is a multiport nonreciprocal optical device. Its function is to make the optical signal can only be transmitted along with the specified port order. The three-port optical circulator we use is transmitted from port1 to port2 and from port2 to port3. The unique parameters in the optical circulator are isolation and directionality. Directionality is also called the crosstalk of the optical circulator. The isolation is due to the imperfection of the optical circulator. When the light is reversed, some light will still pass through. The directionality is caused by the structure of the optical circulator and other reasons, so that part of the light passing through port1 is directly output from port3. The isolation of the optical circulator is defined as the ratio of the input optical power to the output optical power in reverse light transmission \cite{hui2019introduction}. The directionality of the optical circulator is the ratio of the input optical power of port1 to the output optical power of port3 when port2 is terminated and there is no reflection \cite{hui2019introduction}. Therefore, the isolation and directionality of the optical circulator can be expressed as
\begin{equation}
	\begin{aligned}
		I_{21}=10 \lg \left(P_{21} / P_{12}\right), \\
		I_{32}=10 \lg \left(P_{32} / P_{23}\right), \\
		D=10 \lg \left(P_{1 \rm {in}} / P_{3 \rm {out}}\right),
	\end{aligned}
\end{equation}
where $I_{21}$ is the optical isolation from port2 to port1, and $I_{32}$ is the optical isolation from port3 to port2. In the case of port2 to port1, $P_{21}$ is the optical power transmitted by port2, and $P_{12}$ is the optical power received by port1. In the case of port3 to port2, $P_{32}$ is the optical power transmitted by port3, and $P_{23}$ is the optical power received by port2. $D$ is the directionality of optical circulator, where $P_{1 \rm{in}}$ is the optical power input of port1 and $P_{3 \rm{out}}$ is the optical power output of port3.

\indent For different optical structures, the noise introduced by the optical circulator is different. For our scheme, as shown in FIG. \ref{fig6}, the following analysis can be carried out.

\indent First, we analyze the optical circulator of QLT. It receives the light from the laser at port1 and transmits it to QNU at port2. The signal light returned by QNU is then received at port2 and outputs to the detector at port3. As the signal light is returned to the optical circulator, some light will enter port1 from port2 and return to the laser. Therefore, we added an optical isolator behind the laser to avoid damaging the laser and generating noise. Laser usually needs an optical isolator to protect them, so the optical isolator in our scheme is not shown in FIG. \ref{fig6}. For the light received by port1, some light will be directly output to the detector from port3, which will become noise photons and generate noise. The noise can be described as
\begin{equation}
	\varepsilon_{\rm{OC}}^{\rm{QLT}}=\frac{10^{D / 10}V_{\mathrm{A}}}{10^{-\alpha L / 10} 10^{-L_{\mathrm{QNU}} / 10}},
\end{equation}
where $L$ is the length of optical fiber, $\alpha$ is the attenuation coefficient, and the transmittance is $T=10^{-\alpha L / 10}$. Since all QNUs are in parallel, $L_{\mathrm{QNU}}$ is the loss inside one QNU for a round trip. $V_{\mathrm{A}}$ is the modulation variance.

\indent Second, we analyze the optical circulator of QNU. The optical circulator in the QNU receives the light from the QLT at port2, then transmits it to the PM at port3. Afterward, it receives the modulated signal light at port1 and outputs it to the QLT at port2. As the light from the QLT enters the optical circulator, some light enters port1 from port2, which finally enters the PM. However, the light will be screened out by PM due to the unipolarity of PM and will not produce noise. For the signal light received by port1, some light will directly output to the PM from port3, becoming noise photons and generating noise. The noise can be described as
\begin{equation}
	\varepsilon_{\rm{OC}}^{\rm{QNU}}=10^{D / 10}N V_{\mathrm{A}},
\end{equation}
where $N$ is network capacity and also the number of QNU.

\indent In summary, the total noise $\varepsilon_{\rm{OC}}$ introduced by the optical circulator is
\begin{equation}
	\begin{aligned}
		\varepsilon_{\rm{OC}}&=\frac{10^{D / 10} V_{\mathrm{A}}}{10^{-\alpha L / 10} 10^{-L_{\mathrm{QNU}} / 10}}+10^{D / 10}N V_{\mathrm{A}}\\&=10^{D / 10} V_{\mathrm{A}}\left(N+10^{\alpha L / 10} 10^{L_{\mathrm{QNU}} / 10}\right).
	\end{aligned}
\end{equation}

\indent Based on the above analysis, assuming that $D=60\;\rm{dB}$, $\alpha=0.2\;\rm{dB}/km$, $V_{\mathrm{A}}=0.5\;\mathrm{SNU}$, $L_{\mathrm{QNU}}=3\;\mathrm{dB}$, we can get the relation between optical circulator noise and transmission distance, and the relation between optical circulator noise and network capacity, as shown in FIG. \ref{fig7} respectively.

\indent As seen from the figure, the noise of the optical circulator increases gradually with the increase of transmission distance and network capacity, but it is within the acceptable range.

\subsubsection*{d. Other untrusted noise}

\indent In addition to the above mentioned Rayleigh backscattering noise $\varepsilon_{\rm{RB}}$, frequency crosstalk noise $\varepsilon_{\rm{FC}}$ and optical circulator noise $\varepsilon_{\rm{OC}}$, which is the characteristics of the scheme, untrusted noise also includes modulation noise $\varepsilon_{\rm{MO}}$ \cite{laudenbach2018continuous}, amplitude noise $\varepsilon_{\rm{AM}}$ \cite{laudenbach2018continuous} and phase noise $\varepsilon_{\rm{PH}}$ \cite{laudenbach2018continuous}.

\indent The first is modulation noise. Because of the uncertainty of modulation voltage, the noise will be introduced in the modulation process. And we need to find the relationship between the macro uncertainty of voltage and the optical quadrature components. In Ref. \cite{laudenbach2018continuous}, the relationship between the modulation voltage and the optical quadrature components has been derived, which can be expressed as 
\begin{equation}
	\varepsilon_{\rm{MO}}=V_{\mathrm{A}}\left(\pi \frac{\Delta U_{\rm{DAC}}}{U_{\rm{DAC}}}+\frac{1}{2}\left[\pi \frac{\Delta U_{\rm{DAC}}}{U_{\rm{DAC}}}\right]^{2}\right)^{2},
\end{equation}
in which $U_{\rm{DAC}}$ represents the voltage of the digital-to-analog converter in AWG, $\Delta U_{\rm{DAC}}$ denotes the specific deviation of $U_{\rm{DAC}}$, $V_{\mathrm{A}}$ is the modulation variance. In our experiment, $\Delta U_{\mathrm{DAC}}=0.01 U_{\mathrm{DAC}}$ that is decided by the resolution of voltage of AWG, $V_{\mathrm{A}}=0.5 \;\mathrm{SNU}$ that corresponds to our experimental set-up. In this case, $\varepsilon_{\mathrm{MO}}=5.09 \times 10^{-4}$, indicating the contribution of the modulation noise is small in the total noise.

\indent Because the LO used in the coherent detection is physically realized by laser. Thus, the laser has intensity noise and phase noise. The physical cause of the relative intensity noise of the laser is that the laser source uses the principle of excited radiation to produce more photons. In addition, the change in the number of photons caused by the spontaneous radiation will be reflected in the amplitude of the laser, forming amplitude noise. Therefore, the amplitude noise can be calculated as 
\begin{equation}
	\begin{aligned}
	\varepsilon_{\rm{AM}}&=\varepsilon_{\rm{RIN}}^{\rm{sig}}+\varepsilon_{\rm{RIN}}^{\rm{LO}}\\&=V_{\mathrm{A}}\left(\sqrt{\mathrm{RIN}_{\rm{sig}} \Delta v_{\rm{A}}}+0.25 \mathrm{RIN}_\mathrm{LO} \Delta v_{\rm{B}}\right).
\end{aligned}
\end{equation}

\indent Considering that the typical parameter of laser is $\rm{RIN}_{\rm{sig}}=\mathrm{RIN}_{\rm{LO}}=8 \times 10^{-11}\; \mathrm{Hz}^{-1}$, $\Delta v_{\rm{A}}=\Delta v_{\rm{B}}=10\; \mathrm{kHz}$, $V_{\mathrm{A}}=0.5\; \mathrm{SNU}$, amplitude noise $\varepsilon_{\rm{AM}}=4.47 \times 10^{-4}$.

\indent The spontaneous radiation of the laser not only causes the change of intensity but also causes the random change in the frequency of the laser pulse signal, forming phase noise. Due to the spontaneous radiation phenomenon inside the semiconductor laser, the photon generated by it is random in polarization and phase, which will directly affect the amplitude and phase of the light field formed by the excited radiation. For amplitudes, the magnitude is restored to the mean by glazing the radiation field with the inversion of the particle number in the laser medium, but the phase has no such resilience. Therefore, the phase noise can be described as
\begin{equation}
	\varepsilon_{\rm{PH}}=2 \pi \tau V_{\mathrm{A}}\left(\Delta v_{\rm{A}}+\Delta v_{\rm{B}}\right).
\end{equation}

\indent According to the conventional parameter of coherence detection to weak coherent light, assuming that $\tau=1\; \rm{ns}$, $\Delta v_{\rm{A}}=\Delta v_{\rm{B}}=10\; \mathrm{kHz}$, $V_{\mathrm{A}}=0.5\;\mathrm{SNU}$, phase noise $\varepsilon_{\rm{PH}}=6.28 \times 10^{-5}$.

\subsubsection*{e. Trusted noise}

\indent In addition to the untrusted noise mentioned above, there is also trusted noise that can be calibrated or controlled by QLT. Here we briefly describe three kinds of common trusted noise: detector thermal noise $\varepsilon_{\rm{DET}}$, ADC quantization noise $\varepsilon_{\rm{ADC}}$ and common-mode rejection ratio noise $\varepsilon_{\rm{CMRR}}$ \cite{laudenbach2018continuous}.

\indent Detector thermal noise is a kind of white Gaussian noise. Its amplitude distribution is Gaussian, the mathematical expectation is 0, and its power spectral density is constant. Thermal noise is a kind of noise produced by electronic components in the system, mainly by resistors and MOS tubes. The generation of resistance thermal noise is related to the thermal motion of electrons. Therefore, the thermal noise of the detector can be expressed as
\begin{equation}
	\varepsilon_{\rm{DET}}=2 \frac{\mathrm{NEP}^{2} B \tau}{h f P_{\rm{LO}}},
\end{equation}
where $\mathrm{NEP}(\rm{W} / \sqrt{\mathrm{Hz}})$ is the equivalent noise power. $\mathrm{NEP}$ represents the optical signal power required to be input when the signal-to-noise ratio (SNR) is 1. If we take the general parameters $\rm{NEP} = 4.5 \;\rm{pW} / \sqrt{\mathrm{Hz}}$, $B=250\; \mathrm{MHz}$, $\tau=1\; \rm{ns}$, $h f=1.28 \times 10^{-19} \;\mathrm{J}$, $P_{\rm{LO}}=8\; \rm{mW}$, then $\varepsilon_{\rm{DET}}=9.99 \times 10^{-3}$.

\indent The weak coherent state is detected and amplified by a balanced detector such that the final output voltage signal is proportional to the measured canonical component. However, if the output voltage is quantized by an analog-digital converter (ADC), the ADC introduces an additional noise to the weak coherent state, making the excess noise larger. As same as detector noise, we convert a macroscopic physical quantity like ADC noise into noise on the canonical component of the quantum state. Therefore, ADC quantization noise can be expressed as
\begin{equation}
	\varepsilon_{\rm{ADC}}=\frac{2 \tau}{hf(g \rho)^{2} P_{\rm{LO}}}\left(\frac{1}{12} \frac{R_{\rm{U}}^{2}}{2^{2 n}}+V_{\rm{ADC}}\right).
\end{equation}

\indent We can estimate ADC noise by common parameters, $n=10 \;\rm{bit}$, $\tau=1 \;\rm{ns}$, $hf=1.28 \times 10^{19} \;\mathrm{J}$, $g=20 \;\mathrm{k\Omega}$, $\rho=0.85 \; \mathrm{A / W}$, $P_{\rm{LO}}=8 \;\rm{mW}$, $R_{\rm{U}}=1 \;\mathrm{V}$, $V_{\rm{ADC}}=10^{-8} \;\mathrm{V}^{2}$, then $\varepsilon_{\rm{ADC}}=6.05 \times 10^{-4}$.

\indent A practical differential amplifier in a balanced detector not only amplifies the differential currents, but also amplifies their average currents. If a heterodyne detector consisting of two homodyne detectors is used for detection, the final noise introduced by CMRR is
\begin{equation}
	\begin{aligned}
		\varepsilon_{\rm{CMRR}}=\left(\frac{hfV_{\mathrm{A}}^{2}{\rm{RIN}}_{\rm{sig}} \Delta v_{\rm{A}}}{8 \tau P_{\rm{LO}}\left(10^{\rm{CMRR / 10}}\right)^{2}}+\frac{\tau P_{\rm{LO}} {\rm{RIN}}_{\rm{LO}} \Delta v_{\rm{B}}}{2hf \left(10^{\rm{CMRR / 10}}\right)^{2}} \right).
	\end{aligned}
\end{equation}

\indent Assuming that $\rm{CMRR}=30 \;\mathrm{dB}$, $P_{\rm{LO}}=8 \;\mathrm{mW}$, $\tau=1\; \mathrm{ns}$, $V_{\mathrm{A}}=0.5 \;\mathrm{SNU}$, ${\rm{RIN}}_{\rm{sig}}=\rm{RIN}_{LO}=8 \times 10^{-11} \;\mathrm{Hz}^{-1}$, $\Delta v_{\rm{A}}=\Delta v_{\rm{B}}=10 \;\mathrm{kHz}$, $hf=1.28 \times 10^{-19} \;\mathrm{J}$, in this case, $\varepsilon_{\rm{CMRR}}=2.50 \times 10^{-4}$. This noise depends largely on the size of the relative intensity noise of the LO, so the relative intensity noise of the LO is required to be extremely low.

\section*{Conclusion}
\label{sec4}

\indent In conclusion, we propose a flexible and efficient quantum network physical structure, namely RM-QAN. In detail, this quantum access network can make quantum states travel in a circle to transmit the data due to the round-trip structure. It can also support multi-user access through multi-band quantum state transmission and separation. Based on this proposed network, we realize multi-user secure key sharing through CV-QKD. The theoretical noise model of the multi-user CV-QKD has been established, the user capacity and theoretical key rate of the scheme have been discussed, and the proof-of-principle experimental verification has been carried out. The proof-of-principle experiment shows that each QNU can share a practical secret key rate of about $600 \;\rm{bits/s}$ at a transmission distance of $30\;\rm{km}$ with QLT.

\indent Certainly, there are three main challenges in the practical implementation of this scheme: 1. At this stage, we analyze the secret key rate bounds in the asymptotic regime, and the finite-size effects and composable security are not considered, which are critical to the final practical implementation. In such a security framework, we need longer data blocks to overcome the finite-size effect, which requires a higher communication rate and larger data storage, and therefore has requirements for hardware and algorithms. 2. The practical access of more users also needs classical communication protocol related to the data link layer and network layer to realize the information synchronization, such as the handshake protocol between QNU and QLT. Besides, after accessing more QNUs, QLT needs to exchange more classical data in the post-processing stage, requiring the high bandwidth of the classical channel. 3. For practical security, the effective countermeasures of the phase remapping attack \cite{xu2010experimental, xu2020secure} and the Trojan-Horse attack \cite{gisin2006trojan} have been proposed and can be added in further implementation.

\indent The advantage of our scheme is the simple optical structure in the physical layer. Specifically, the round-trip structure only requires one laser and one detector in the entire network, and only one modulator and one circulator need to be plugged in when a new user accesses. Performance evaluation shows such networks have a low physical excess noise for each user theoretically and can support multi-user access and quantum secure key generation. Moreover, this scheme can coexist with classical communication, since classical communication can adopt different frequency bands in FDM. Combined with the comprehensive protocols of the data link layer and network layer and practical security countermeasures, this scheme can be an effective solution for the QKD access network establishment. This work lays the foundation for the subsequent establishment and application of a large-scale and multi-user quantum access network.

\section*{Appendix}

In the following, we derive the secret key rate of DMCS CV-QKD under collective attack. Specifically, in the case of reverse negotiation, the secret key rate for unit system repetition rate can be written as
\begin{equation}
K_{p}=\beta I_{\mathrm{AB}}-\chi_{\mathrm{BE}},
\end{equation}
where $\beta \in(0,1)$ is reverse negotiation efficiency, $I_{\mathrm{AB}}$ is the mutual information between Alice and Bob, and $\chi_{\mathrm{BE}}$ is the maximum information that Eve can extract from Bob's secret key. According to Bob's measurement variance $V_{\mathrm{B}}=\eta T\left(V+\chi_{\rm {tot}}\right)$ and conditional variance $V_{\mathrm{B} \mid \mathrm{A}}=\eta T\left(1+\chi_{\mathrm{tot}}\right)$, where $T$ is the transmittance of channel, $I_{\mathrm{AB}}$ can be calculated as
\begin{equation}
\begin{aligned}
I_{\mathrm{AB}}^{\mathrm{hom}} &= \frac{1}{2} \log _{2} \frac{V_{\mathrm{B}}}{V_{\mathrm{B} \mid \mathrm{A}}}=\frac{1}{2} \log _{2} \frac{V+\chi_{\rm {tot}}}{1+\chi_{\rm {tot}}}, \\
I_{\mathrm{AB}}^{\mathrm{het}} &= 2 \times \frac{1}{2} \log _{2} \frac{V_{\mathrm{B}}}{V_{\mathrm{B} \mid \mathrm{A}}}=\log _{2} \frac{V+\chi_{\rm {tot}}}{1+\chi_{\rm {tot}}},
\end{aligned}
\end{equation}
where $V=V_{\mathrm{A}}+1$ is the equivalent variance of pure two-mode entangled states. $V_{\mathrm{A}}=2 \alpha^{2}$ is the modulation variance in the preparation-measure model. Because heterodyne detection measures two quadrature components at the same time, the mutual information is multiplied by the coefficient $2$. The total noise $\chi_{\rm {tot}}$ can be described as
\begin{equation}
\begin{aligned}
\chi_{\rm {tot}} &= \chi_{\rm {line}}+\chi_{\rm {det}}/ T, \\
\chi_{\rm {line}} &=1 / T-1+\varepsilon,
\end{aligned}
\end{equation}
where $\chi_{\rm {line}}$ denotes the channel noise, $\chi_{\rm {det}}$ represents the detection noise, $\varepsilon$ is the excess noise. For homodyne detection, $\chi_{\rm {det}}=\chi_{\mathrm{hom}}=\left[(1-\eta)+v_{e l}\right] / \eta$, and for heterodyne detection $\chi_{\rm {det}}=\chi_{\mathrm{het}}=\left[1+(1-\eta)+2 v_{e l}\right] / \eta$. $\eta$ denotes quantum efficiency, $v_{e l}$ represents electrical noise.

The core of secret key rate calculation is to evaluate the upper bound of the information Eve steals. Under collective attack, the Holevo bound is used to limit the maximum information Eve can extract from Bob, so $\chi_{\rm{BE}}$ is
\begin{equation}
\chi_{\mathrm{BE}}=S\left(\rho_{\mathrm{E}}\right)-\int d m_{\mathrm{B}} p\left(m_{\mathrm{B}}\right) S\left(\rho_{\mathrm{E}}^{m_{\mathrm{B}}}\right),
\end{equation}
where $m_{\mathrm{B}}$ represents the measurements of Bob, and $p\left(m_{\mathrm{B}}\right)$ represents the probability density of the measurements, $\rho_{\mathrm{E}}^{m_{\mathrm{B}}}$ is Eve's conditional quantum state under Bob's measurements, and $S$ denotes the Von-Neumann entropy of quantum state $\rho$. Because Eve's system can purify the system $\mathrm{AB}_{1}$, Bob's measurement can purify system AEFG, and $S\left(\rho_{\mathrm{AFG}}^{m_{\mathrm{B}}}\right)$ and $m_{\mathrm{B}}$ are independent of each other in the protocol, so $\chi_{\mathrm{BE}}$ can be simplified as
\begin{equation}
\chi_{\mathrm{BE}}=S\left(\rho_{\mathrm{AB}_{1}}\right)-S\left(\rho_{\mathrm{AFG}}^{m_{\mathrm{B}}}\right).
\end{equation}

Theoretical security analysis of CV-QKD protocol under collective attack shows that under the condition of known covariance matrix $\gamma_{\mathrm{AB}_{1}}$ of state $\rho_{\mathrm{AB}_{1}}$, if Eve's eavesdropping operation is a Gaussian operation, it can get the most information, which is called "Gaussian attack optimality theorem". The theorem states that if the final quantum state $\rho_{\mathrm{AB}_{1}}$ shared by Alice and Bob is regarded as a Gaussian state, the calculated stolen information by Eve is an upper bound of the real stolen information. The information entropy calculation of the Gaussian state is relatively simple, which makes the above equation can be simplified as
\begin{equation}
\chi_{\mathrm{BE}}=\sum_{i=1}^{2} \mathrm{G}\left(\frac{\lambda_{i}-1}{2}\right)-\sum_{i=3}^{5} \mathrm{G}\left(\frac{\lambda_{i}-1}{2}\right),
\end{equation}
where $\mathrm{G}(x)=(x+1) \log _{2}(x+1)-x \log _{2}x$. $\lambda_{i}$ is the symplectic eigenvalue of the covariance matrix, where $\lambda_{1,2}$ corresponds to the covariance matrix $\gamma_{\mathrm{AB}_{1}}$ of representational state $\rho_{\mathrm{AB}_{1}}$ and $\lambda_{3,4,5}$ corresponds to the covariance matrix $\gamma_{\mathrm{AFG}}^{m_{\mathrm{B}}}$ of representational state $\rho_{\mathrm{AFG}}^{m_{\mathrm{B}}}$. On the one hand, the covariance matrix $\gamma_{\mathrm{AB}_{1}}$ only depends on the Alice and the channel, which is independent of the specific detection mode. It can be expressed as
\begin{equation}
\gamma_{\mathrm{AB}_{1}}=\left[\begin{array}{cc}
V \cdot \mathrm{I}_{2} & \sqrt{T} Z_{4} \cdot \sigma_{z} \\
\sqrt{T} Z_{4} \cdot \sigma_{z} & T\left(V+\chi_{\rm {line}}\right) \cdot \mathrm{I}_{2}
\end{array}\right],
\label{fu26}
\end{equation}
where $\mathrm{I}_{2}=\operatorname{diag}(1,1)$, 
$\sigma_{z}=\operatorname{diag}(1,-1)$. $Z_{4}$ reflects the correlation between patterns $\mathrm{AB}_{1}$. It can be described as
\begin{equation}
Z_{4}=2 \alpha^{2} \left(l_{0}^{3 / 2} l_{1}^{-1 / 2}+l_{1}^{3 / 2} l_{2}^{-1 / 2}+l_{2}^{3 / 2} l_{3}^{-1 / 2}+l_{3}^{3 / 2} l_{0}^{1 / 2}\right),
\end{equation}
where
\begin{equation}
\begin{aligned}
l_{0,2}&=\frac{1}{2} e^{-\alpha^{2}}\left[\cosh \alpha^{2} \pm \cos \alpha^{2}\right], \\
l_{1,3}&=\frac{1}{2} e^{-\alpha^{2}}\left[\sinh \alpha^{2} \pm \sin \alpha^{2}\right].
\end{aligned}
\end{equation}

It can be found that this matrix is similar to the covariance matrix in Gaussian modulation coherent state (GMCS) CV-QKD protocol, except that the Einstein–Podolsky–Rosen (EPR) correlation in GMCS CV-QKD protocol is $Z_{\rm{G}}=\sqrt{\left(V^{2}-1\right)}$. When $V_{\mathrm{A}}<0.5$, $Z_{4}$ is extremely close to $Z_{\rm{G}}$. Under this condition, it can be considered that the information $\chi_{\rm{BE}}$ Eve steals from Bob is equal in both protocols. Based on this conclusion, the secret key rate can be deduced according to the GMCS CV-QKD protocol, and the corresponding parameter $A$ and $B$ are
\begin{equation}
\begin{aligned}
&A=V^{2}+T^{2}\left(V+\chi_{\rm {line}}\right)^{2}-2 T Z_{4}^{2}, \\
&B=\left(T V^{2}+T V \chi_{\rm {line}}-T Z_{4}^{2}\right)^{2}.
\end{aligned}
\end{equation}

On the other hand, matrix $\gamma_{\mathrm{AFG}}^{m_{\mathrm{B}}}$ can be calculated as
\begin{equation}
\gamma_{\mathrm{AFG}}^{m_{\mathrm{B}}}=\gamma_{\mathrm{AFG}}-\sigma_{\mathrm{AFGB}_{3}}^{T} H \sigma_{\mathrm{AFGB}_{3}},
\end{equation}
where the symplectic matrix $H$ represents the measurement method in the pattern $\mathrm{B}_{3}$. For homodyne detection, $H_{\mathrm{hom}}=\left(X \gamma_{\mathrm{B}_{3}} X\right)^{\mathrm{MP}}$, where $X=\operatorname{diag}(1,0)$. MP stands for Moore-Penrose inverse of the matrix. For heterodyne detection, $H_{\rm {het}}=\left(\gamma_{B_{3}}+I_{2}\right)^{-1}$. The matrix $\gamma_{B_{3}}, \gamma_{\mathrm{AFG}}$ and $\sigma_{\mathrm{AFGB}_{3}}$ can be obtained by decomposing the following covariance matrix
\begin{equation}
\gamma_{\mathrm{AFGB}_{3}}=\left[\begin{array}{cc}
\gamma_{\mathrm{AFG}} & \sigma_{\mathrm{AFGB}_{3}}^{T} \\
\sigma_{\mathrm{AFGB}_{3}} & \gamma_{\mathrm{B}_{3}}
\end{array}\right],
\end{equation}
which can be obtained by the transformation of the matrix describing the system $\mathrm{AB}_{3} \mathrm{FG}$. The matrix of the system $\mathrm{AB}_{3} \mathrm{FG}$ is
\begin{equation}
	\gamma_{\mathrm{AB}_{3} \mathrm{FG}}=\left(Y^{\mathrm{BS}}\right)^{T}\left[\mathrm{\gamma}_{\mathrm{AB}_{1}} \oplus \gamma_{\mathrm{F}_{0} \mathrm{G}}\right] Y^{\mathrm{BS}},
\end{equation}
where $\gamma_{\mathrm{AB}_{1}}$ is Given in Eq. \ref{fu26}. $\gamma_{\mathrm{F}_{0} \mathrm{G}}$ describes the EPR state of variance $v$, which is used for equivalent electrical noise of the detector. So $\gamma_{\mathrm{F}_{0} \mathrm{G}}$ is
\begin{equation}
\gamma_{\mathrm{F}_{0} \mathrm{G}}=\left[\begin{array}{cc}
v \cdot \mathrm{I}_{2}  & \sqrt{\left(v^{2}-1\right)} \cdot \sigma_{z} \\
\sqrt{\left(v^{2}-1\right)} \cdot \sigma_{z} & v \cdot \mathrm{I}_{2}
\end{array}\right],
\end{equation}
where $v$ depends on the detection method. For homodyne detection, $v=\eta \chi_{\mathrm{hom}} /(1-\eta)=1+v_{e l} /(1-\eta)$, and for heterodyne detection $v=\left(\eta \chi_{\mathrm{het}}-1\right) /(1-\eta)=1+2 v_{e l} /(1-\eta)$. Finally, the matrix $Y^{\mathrm{BS}}$ describes the function of the BS on pattern $\rm B_{2}$ and pattern $\rm F_{0}$, which is used for the equivalent quantum efficiency of the detector. Therefore, the matrix $Y^{\mathrm{BS}}$ can be denoted as
\begin{equation}
\begin{aligned}
Y_{\mathrm{B}_{2} \mathrm{F}_{\mathrm{0}}}^{\mathrm{BS}} &=\left[\begin{array}{cc}
\sqrt{\eta} \cdot \mathrm{I}_{2} & \sqrt{1-\eta} \cdot \mathrm{I}_{2} \\
-\sqrt{1-\eta} \cdot \mathrm{I}_{2} & \sqrt{\eta} \cdot \mathrm{I}_{2}
\end{array}\right], \\
Y^{\mathrm{BS}} &=\mathrm{I}_{\mathrm{A}} \oplus Y_{\mathrm{B}_{2} \mathrm{F}_{\mathrm{0}}}^{\mathrm{BS}} \oplus \mathrm{I}_{\mathrm{G}}.
\end{aligned}
\end{equation}

After obtaining the above matrix, we can find the symplectic eigenvalues of the matrix $\gamma_{\mathrm{AFG}}^{m_{\mathrm{B}}}$. In the following, we directly give their calculation formula
\begin{equation}
\lambda_{3,4}^{2}=\frac{1}{2}\left(C \pm \sqrt{C^{2}-4 D}\right),
\end{equation}
where $C, D$ are determined by the specific detection method. For the homodyne detection,
\begin{equation}
\begin{aligned}
C_{\mathrm{hom}}&=\frac{V \sqrt{B}+T\left(V+\chi_{\rm {line}}\right)+A \chi_{\rm {hom }}}{T\left(V+\chi_{\rm {tot}}\right)}, \\
D_{\mathrm{hom}}&=\sqrt{B} \frac{V+\sqrt{B} \chi_{\rm {hom }}}{T\left(V+\chi_{\rm {tot}}\right)},
\end{aligned}
\end{equation}
and for the heterodyne detection,
\begin{equation}
\begin{aligned}
C_{\rm {het}} &=\frac{1}{\left(T\left(V+\chi_{\rm {tot}}\right)\right)^{2}}\left[A \chi_{\rm {het}}^{2}+B+1+2 \chi_{\rm {het}}\right.  \\ \phantom{=\;\;} &\left. \left(V \sqrt{B}+T\left(V+\chi_{\rm {line}}\right)\right)+2 T\left(V^{2}-1\right)\right], \\
D_{\rm {het}} &=\left(\frac{V+\sqrt{B} \chi_{\rm {het}}}{T\left(V+\chi_{\rm {tot}}\right)}\right)^{2}.
\end{aligned}
\end{equation}

And the last symplectic eigenvalue is
\begin{equation}
\begin{aligned}
\lambda_{5}=1.
\end{aligned}
\end{equation}

\begin{backmatter}

\bmsection{Funding}
National Key Research and Development Program of China (2016YFA0302600), National Natural Science Foundation of China (62101320, 61671287, 61971276), Shanghai Municipal Science and Technology Major Project (2019SHZDZX01), and the Key R\&D Program of Guangdong province (2020B030304002).

\bmsection{Author contributions}
G. Z. conceived the research. Y. X. and T. W. carried out the experiment. Y. X., T. W. and H. Z analyzed the data and wrote the manuscript. P. H. provided the technical guide for secret key rate analysis and post processing. All authors contributed to the data collection, discussed the results and reviewed the manuscript.
		
\bmsection{Disclosures} 
The authors declare no conflicts of interest.
		
\bmsection{Data Availability} 
The data that support the findings of this study are available from the corresponding author upon reasonable request.
		
\end{backmatter}

\end{document}